\documentclass[12pt,draftclsnofoot,onecolumn]{IEEEtran}
\usepackage{amsfonts}
\usepackage{amsfonts}
\usepackage{setspace}

\usepackage{enumerate}

\usepackage{amsmath}
\usepackage{amssymb}
\usepackage[dvips]{graphicx}
\usepackage{epsfig}
\usepackage{hyperref}

\usepackage{subfigure}

\newcommand{\tsnr}{{\text{\footnotesize{SNR}}}}
\newcommand{\tmin}{\text{min}}
\newcommand{\tmax}{\text{{max}}}
\newcommand{\E}{\mathbb{E}}

\newcommand{\Pb}{\bar{P}}

\newcommand{\figsize}{0.45}
\newcommand{\subfigsize}{0.45}

\newcommand{\bR}{{\mathbf{R}}}
\newcommand{\bD}{{\mathbf{D}}}
\newcommand{\bS}{{\mathbf{S}}}

\newcommand{\btheta}{\bar{\theta}}
\newcommand{\bbtheta}{\underline{\theta}}
\newcommand{\uutheta}{\check{\theta}}
\newcommand{\vvtheta}{\overset{\circ}{\theta}}

\newtheorem{Lem1}{Proposition}
\newtheorem{Lem}{Theorem}
\newtheorem{Lemm}{Lemma}
\newtheorem{Rem}{Remark}

\newtheorem{Def}{Definition}
\newtheorem{Assu}{Assumption}

\begin{document}

%
\title{Statistical Delay Tradeoffs in Buffer-Aided Two-Hop Wireless Communication Systems}



%
\author{
\vspace{1cm}
\authorblockN{Deli Qiao and M. Cenk Gursoy}
\thanks{D. Qiao is with the School of Information Science and Technology, East China Normal University, Shanghai, China, 200241 (e-mail: dlqiao@ce.ecnu.edu.cn). M. Cenk Gursoy is with the Department of Electrical Engineering and Computer Science, Syracuse University, Syracuse, NY 13244 (email: mcgursoy@syr.edu)}
\thanks{This work was supported in part by the National Natural Science Foundation of China under Grants (61571191, 61572192). The material in
this paper has been presented in part at the 2015 IEEE Global Communications Conference (Globecom), San Diego, United States, Dec 2015. }}


\maketitle

\begin{abstract}
This paper analyzes the impact of statistical delay constraints on the achievable rate of a two-hop wireless communication link, in which the communication between a source and a destination is accomplished via a buffer-aided relay node. It is assumed that there is no direct link between the source and the destination, and the buffer-aided relay forwards the information to the destination by employing the decode-and-forward scheme. Given statistical delay constraints specified via maximum delay and delay violation probability, the tradeoff between the statistical delay constraints imposed on any two concatenated queues is identified. With this characterization, the maximum constant arrival rates that can be supported by this two-hop link are obtained by determining the effective capacity of such links as a function of the statistical delay constraints, signal-to-noise ratios (\tsnr) at the source and relay, and the fading distributions of the links. It is shown that asymmetric statistical delay constraints at the buffers of the source and relay node can improve the achievable rate. Overall, the impact of the statistical delay tradeoff on the achievable throughput is provided.
\end{abstract}

\begin{IEEEkeywords}
Two-hop wireless links, statistical delay constraints, quality of service (QoS) constraints, fading channels, effective capacity, delay violation probability, full-duplex relaying.
\end{IEEEkeywords}

\begin{spacing}{1.53}
\section{Introduction}

With the widespread use of smart-phones and tablets, the volume of global mobile traffic has increased explosively in recent years. The portion of multimedia data, such as mobile video and voice over IP (VoIP),  has surged significantly within this wireless traffic \cite{cisco}. In such multimedia traffic, delay is an important consideration. Meanwhile, providing deterministic quality of service (QoS) guarantees is challenging in wireless systems, since the instantaneous rate of the channel varies randomly depending on numerous factors, such as mobility, changing environment and multipath fading \cite{book}. Therefore, providing statistical QoS guarantees is more suitable in such randomly-varying wireless environment. 

Effective bandwidth theory has been developed to analyze high-speed communication systems operating under statistical queueing constraints \cite{chang1}, \cite{kelly}. The queueing constraints are imposed on buffer violation probabilities and are
specified by the QoS exponent $\theta$, which dictates the exponential decay rate of the queue length in the stable state.
Also, Chang and Zajic have characterized the effective bandwidths of time-varying departure processes in \cite{chang2}, which can be utilized to analyze the volatile wireless systems. Moreover, Wu and Negi in
\cite{dapeng} defined the dual concept of effective capacity, which
provides the maximum constant arrival rate that can be supported by
a given departure process while satisfying statistical delay
constraints. The analysis and application of effective capacity in various
settings have attracted much interest recently (see e.g. \cite{tangzhangcross2}-\cite{qingyang} and references therein). 

In this paper, we study the achievable rate of two-hop systems operating under statistical delay constraints. In particular, we assume that there are buffers at both the source and the relay nodes, and consider the queueing delay introduced by the buffers. Note that \cite{tangrelay}-\cite{qingyang} have also recently investigated the effective capacity of the relay channels. For instance, Tang and Zhang in \cite{tangrelay} analyzed the power allocation policies of relay networks, where the relay node is assumed to have no queue, i.e., the packets arriving to the relay node are forwarded immediately. In \cite{liu-cooperation}, Liu \emph{et al.} considered the cooperation of two users for data transmission, where the interchanged data goes through only the queue of the other user. Parag and Chamberland in \cite{butterfly} provided a queueing analysis of a butterfly network with constant rate for each link, while assuming that there
is no congestion at the intermediate nodes. The effective capacity of the two-hop link in the presence of the statistical queueing constraints at the source and relay node is given in \cite{deli-twohop}, and the performance for multi-relay links is analyzed in \cite{dualhop}.

In this work, as a significant departure from previous works, we consider statistical end-to-end delay constraints, imposed as the limitations on the maximum delay and delay violation probability. Note that statistical end-to-end delay analysis can also be found in \cite{wurelay}-\cite{qingyang}. In \cite{wurelay}, Wu and Negi considered statistical end-to-end delay constraints for half-duplex relays, and gave an effective capacity formulation with time allocation to the different hops. In \cite{khalekrelay}-\cite{qingyang}, the authors considered the statistical end-to-end delay constraints of multi-hop links, while assuming that the statistical delay violation probability of the queues are equal. However, it is possible that the relay can tolerate more stringent delay constraints while not affecting the system performance \cite{deli-twohop}. Therefore, we seek to determine the optimal statistical QoS exponents of the buffers under given \emph{end-to-end} delay constraints. Additionally, we note that the analysis of buffer-aided systems have attracted much interest recently (see e.g., \cite{bufferrelay}-\cite{directrelay} and reference therein). In such analysis, the authors considered the case that only the relay node has buffer, and the average queueing delay is investigated \cite{bufferaid}. The contributions can be summarized as follows:
 \begin{enumerate}
 \item We characterize the tradeoff between the statistical delay constraints at the source and relay nodes, providing a framework for dynamically adjusting the delay constraints of any two interacting queues.
 \item With the identified interplay, we then derive the effective capacity of the two-hop links under a target statistical end-to-end delay constraint by optimizing over the statistical queueing constraints at the queues of the source and relay nodes.
 \item We also describe a method for obtaining the effective capacity in such settings. Additionally, we show that symmetric delay constraints at the two buffers do not always lead to the optimal performance. Instead, asymmetric delay constraints, e.g., when the delay constraint at one queue is more relaxed, can lead to larger achievable rates for the two-hop system, which we verify via numerical results. Moreover, it is demonstrated that the improvement is affected by the statistical delay constraints, the signal-to-noise ratio (SNR) levels and the channel conditions of the links.
     \end{enumerate}

The rest of this paper is organized as follows. In Section II, the
system model and necessary preliminaries are described. In Section
III, we present the tradeoff between the statistical delay constraints of any two concatenated queues. We describe our main results for block-fading channels in Section IV, with numerical results provided in Section V. Finally, in Section VI, we conclude the paper.

\section{Preliminaries}

\subsection{System Model}

\begin{figure}
\begin{center}
\includegraphics[width=\figsize\textwidth]{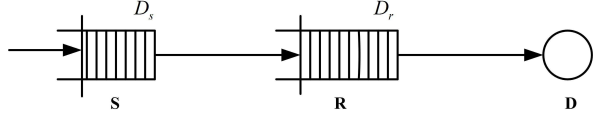}
\caption{The system model.}\label{fig:systemmodel}
\end{center}
\end{figure}

The two-hop communication link is depicted in Figure
\ref{fig:systemmodel}. In this model, source $\bS$ is sending
information to the destination $\bD$ with the help of the intermediate relay
node $\bR$. We assume that there is no direct link between $\bS$ and $\bD$ (which, for instance, holds, if these nodes are sufficiently far apart in distance). Both the source and the intermediate relay nodes are equipped with buffers. Hence, for the information flow of such links, the queueing delay experienced is given by
$D = D_s+D_r$,
where $D_s$ and $D_r$ denote the stationary delay experienced in the queue at the source and relay node, respectively.





We consider a full-duplex relay, and hence assume that reception and transmission can be performed simultaneously at the relay node. Note that full-duplex relaying can be achieved through some form of analog self-interference cancellation followed by digital self-interference cancellation in the baseband domain \cite{fd-1}, \cite{fd-2}. In the $i$th symbol
duration, the signal $Y_r$ received at the relay from the source and the signal $Y_d$ received at the destination from the relay can be expressed as
\begin{align}
Y_r[i]&=g_1[i]X_1[i]+n_1[i], \\
Y_d[i]&=g_2[i] X_2[i]+n_2[i],
\end{align}
where $X_j$ for $j=\{1, 2\}$ denote the inputs for the links $\bS-\bR$ and
$\bR-\bD$, respectively. More specifically, $X_1$ is the signal sent from the source and $X_2$ is sent from the relay. The inputs are subject to individual average
energy constraints $\E\{|X_j|^2\}\le \Pb_j/B, j=\{1,2\}$ where $B$
is the bandwidth. Assuming that the symbol rate is $B$
complex symbols per second, we can easily see that the symbol energy
constraint of $\Pb_j/B$ implies that the channel input has a power
constraint of $\Pb_j$. We assume that the fading coefficients $g_{j},
j=\{1,2\}$ are jointly stationary and ergodic discrete-time
processes, and we denote the magnitude-square of the fading
coefficients by $z_j[i]=|g_j[i]|^2$.  Above, in the channel input-output
relationships, the noise component $n_j[i]$ is a zero-mean,
circularly symmetric, complex Gaussian random variable with variance
$\E\{|n_j[i]|^2\} = N_j$ for $j = 1,2$. The additive Gaussian noise
samples $\{n_j[i]\}$ are assumed to form an independent and
identically distributed (i.i.d.) sequence. We denote the
signal-to-noise ratios as $\tsnr_j=\frac{\Pb_j}{N_j B}$.


\subsection{Statistical Delay Constraints}

Suppose that the queue is stable and there exists a unique $\theta>0$ such that
\begin{align}\label{eq:QoSstabdef}
\Lambda_A(\theta)+\Lambda_C(-\theta)=0,
\end{align}
where $\Lambda_A(\theta)$ and $\Lambda_C(\theta)$ are the logarithmic moment generating functions (LMGFs) of the arrival and service processes, respectively.
Then, \cite{chang2}
\begin{small}
\begin{align} \label{eq:QoSexponentdef}
\lim_{Q_{\tmax}\to\infty}\frac{\log
\Pr\{Q>Q_{\tmax}\}}{Q_{\tmax}}=-\theta.
\end{align}
\end{small}
where $Q$ is the stationary queue length. Throughout the text, logarithm expressed without a base, i.e., $\log(\cdot)$, refers to
the natural logarithm $\log_e(\cdot)$.

We need to guarantee that the statistical delay performance of the two-hop link is not worse than the statistical delay performance specified by $(\varepsilon,D_\tmax)$, where $\varepsilon$ is the limitation on the statistical delay violation probability, and $D_\tmax$ is the maximum tolerable delay. Note that the end-to-end delay consists of the queueing and transmission delays. As indicated in \cite[Section IV]{itcn}, the flow of data bits are treated as the flow of a fluid in the theory of effective bandwidth, in which case the transmission delay can be negligible if $T\ll D_\tmax$. The end-to-end delay can be approximated by the queueing end-to-end delay \cite{tangzhangcross2}, \cite{heathqos}. Assume that the first-in first-out (FIFO) queues are saturated, and hence they always attempt to transmit \cite{sdrrelay}. Then, the queueing delay violation probability can be written equivalently as \cite{tangzhangcross2}, \cite{heathqos}
\begin{small}
\begin{align}\label{eq:sddelay}
\Pr\{D>D_{\tmax}\} \doteq  e^{-J(\theta) D_\tmax}
\end{align}
\end{small}
where we define $f(x)\doteq e^{cx}$ when $\lim_{x\to\infty}\frac{\log f(x)}{x}=c$, and
\begin{align}
J(\theta) = \theta \delta = -\Lambda_C(-\theta)
\end{align}
is the statistical delay exponent associated with the queue, with $\Lambda_C(\theta)$ denoting the LMGF of the service rate, and $\delta$ is decided by the arrival and departure processes jointly. Note that the larger $J(\theta)$, the smaller the delay violation probability is, implying more stringent delay constraints. Now, we can express the probability density function of the random variable $D$ as
\begin{align}
p_D(x) = \frac{\partial}{\partial x} \left(1-\Pr\{D>x\}\right) \doteq J(\theta) e^{-J(\theta) x}.
\end{align}

Consider the two concatenated queues as depicted in Fig. \ref{fig:systemmodel}. For the queueing constraints specified by $\theta_1$ and $\theta_2$ with (\ref{eq:QoSstabdef}) satisfied for each queue, we define
\begin{align}\label{eq:J1J2eq}
J_1(\theta_1)=-\Lambda_{C,1}(-\theta_1),\,\,\text{and}\,\, J_2(\theta_2)=-\Lambda_{C,2}(-\theta_2),
\end{align}
where $\Lambda_{C,1}(\theta_1)$ and $\Lambda_{C,2}(\theta_1)$ are the LMGFs of the service rates of queues at the source and relay nodes, respectively. In the two-hop system, we can express the end-to-end delay violation probability as
\begin{small}
\begin{align}
\Pr\{D_1+D_2>D_\tmax\} &= 1 - \int_0^{D_\tmax} \int_{0}^{D_\tmax - D_1}p_{D}(D_1)p_D(D_2)dD_2dD_1\\
& \doteq \left\{
\begin{array}{ll}
\frac{J_1(\theta_1)e^{-J_2(\theta_2)D_{\tmax}}-J_2(\theta_2)e^{-J_1(\theta_1)D_{\tmax}}}{J_1(\theta_1)-J_2(\theta_2)},& J_1(\theta_1)\neq J_2(\theta_2),\\
\left(1+J_1(\theta_1)D_{\tmax}\right)e^{-J_1(\theta_1)D_{\tmax}},&J_1(\theta_1)=J_2(\theta_2).
\end{array}\right.\label{eq:delayprob}
\end{align}
\end{small}
Note that we should satisfy
\begin{align}\label{eq:queue12cond}
\Pr\{D_1+D_2>D_{\tmax}\}\le \varepsilon.
\end{align}

\subsection{Effective Capacity}

We can dynamically control the delay constraints at the queues of the source and relay nodes specified by $J_1(\theta_1)$ and $J_2(\theta_2)$ as long as the statistical end-to-end delay performance (\ref{eq:queue12cond}) can be guaranteed. At the same time, for each realization of $(\theta_1,\theta_2)$, assume that the constant arrival rate at the source is $ R\ge0$, and the
channels operate at their capacities. To satisfy the queueing constraint
at the source, we must have
\begin{align}\label{eq:cond1}
\tilde{\theta}\ge \theta_1,
\end{align}
where $\tilde{\theta}$ is the solution to
\begin{align}\label{eq:cond1equ}
R=-\frac{\Lambda_{sr}(-\tilde{\theta})}{\tilde{\theta}},
\end{align}
and $\Lambda_{sr}(\theta)$ is the LMGF of the instantaneous capacity of the $\bS-\bR$
link.

In order to satisfy the queueing constraint of the intermediate relay
node $\bR$, we must have
\begin{align}\label{eq:cond2}
\hat{\theta}\ge\theta_2,
\end{align}
where $\hat{\theta}$ is the solution to
\begin{align}\label{eq:cond2equ}
\Lambda_r(\hat{\theta})+\Lambda_{rd}(-\hat{\theta})=0.
\end{align}
Above, $\Lambda_r(\theta)$ is the LMGF of the arrival process to the queue at the relay, and $\Lambda_{rd}(\theta)$ is the LMGF of the instantaneous capacity of the $\bR-\bD$
link.

Note that we can obtain the effective capacity $R_E(\theta_1,\theta_2)$ with $(\theta_1,\theta_2)$ following the method provided in \cite[Theorem 2]{deli-twohop} (Appendix \ref{app:prev}).\footnote{We include the theorem in Appendix \ref{app:prev} for the reader's convenience.} Denote $\Omega$ as the set of pairs $(\theta_1,\theta_2)$ such that (\ref{eq:queue12cond}) can be satisfied. After these characterizations, effective capacity of the two-hop
communication model under statistical delay constraints $(\varepsilon,D_\tmax)$ can be formulated as follows.
\begin{Def}\label{def:ecdef}
The effective capacity of the two-hop communication link with statistical delay constraints specified by $(\varepsilon,D_\tmax)$ is given by
\begin{align}\label{eq:effdefi}
R_\varepsilon(\varepsilon,D_\tmax)=\sup_{(\theta_1,\theta_2)\in\Omega} R_E(\theta_1,\theta_2)
\end{align}
where $\Omega$ is the set of all feasible $(\theta_1, \theta_2)$ satisfying (\ref{eq:queue12cond}). Hence, effective capacity is now the maximum constant arrival rate that can be supported by the two-hop channels under the end-to-end statistical delay constraints.
\end{Def}

\section{Statistical Delay Tradeoffs}

For the following analysis, we first characterize the relation between $J_1(\theta_1)$ and the associated minimum $J_2(\theta_2)$ satisfying the statistical delay constraint (\ref{eq:queue12cond}). We have the following results.
\begin{Lemm}\label{lemm:J1J2relation}
Consider the following function
\begin{small}
\begin{align}\label{eq:J1J2function}
\vartheta(J_1(\theta_1),J_2(\theta_2))=\frac{J_2(\theta_2)e^{-J_1(\theta_1)D_\tmax} - J_1(\theta_1)e^{-J_2(\theta_2)D_\tmax}}{J_2(\theta_2)-J_1(\theta_1)} = e^{-J_0D_\tmax}=\varepsilon, \,\text{for} \, 0\le\varepsilon\le1,
\end{align}
\end{small}
where $J_0=-\frac{\log\varepsilon}{D_\tmax}$ is defined as the statistical delay exponent associated with $(\varepsilon,D_\tmax)$. Denoting $J_2(\theta_2) = \Phi(J_1(\theta_1))$ as a function of $J_1(\theta_1)$, we have the following properties:
\begin{enumerate}[a)]

\item $\Phi(J_1(\theta_1))$ is continuous. Moreover, for $J_1(\theta_1)=J_{th}(\varepsilon)$, we have
\begin{align}
\Phi(J_1(\theta_1)) =  J_{th}(\varepsilon)
\end{align}
where
\begin{align}\label{eq:Jfunctioncond}
J_{th}(\varepsilon) = -\frac{1}{D_{\tmax}}\left(1+\mathcal{W}_{-1}\left(-\frac{\varepsilon}{e}\right)\right),
\end{align}
with $\mathcal{W}_{-1}(\cdot)$ denoting the Lambert W function, which is the inverse function of $y=xe^x$ in the range $(-\infty,-1]$.

\item $\Phi$ is strictly decreasing in $J_1(\theta_1)$.

\item $\Phi$ is convex in $J_1(\theta_1)$.

\item $J_1(\theta_1)\in[J_0,\infty)$, and $J_2(\theta_2)=\Phi(J_1(\theta_1))\in[J_0,\infty)$.

\end{enumerate}
\end{Lemm}
\emph{Proof: }See Appendix \ref{app:J1J2relation}.
\begin{figure}
\begin{center}
\includegraphics[width=\figsize\textwidth]{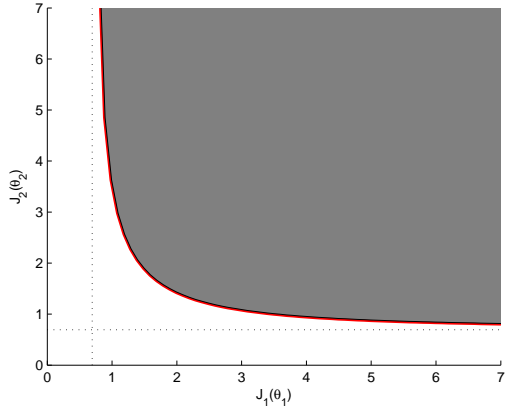}
\caption{$J_2$ v.s. $J_1$. $D_\tmax=1$ sec. $\varepsilon=0.05$.}\label{fig:delaybound=05_dmax=1}
\end{center}
\end{figure}

\begin{Rem}
The above properties can be understood intuitively. Larger $J_1(\theta_1)$ enforces more stringent delay constraints on queue 1 (i.e., the source queue), and we can loosen the delay constraints for the queue 2 (i.e., the relay queue), and vice versa. When either queue is subject to a deterministic constraint, i.e., $\theta=\infty$, the delay violation occurs only at the other queue. In Fig. \ref{fig:delaybound=05_dmax=1}, we plot $J_2$ as a function of $J_1$ for the case with $\varepsilon=0.05$ and $D_\tmax=1$ sec for illustration. Note that only $(J_1,J_2)$ in the dark region are feasible to achieve the statistical delay performance. As can be seen from the figure, the curve given by the lower boundary matches the properties in the lemma.
\end{Rem}

\section{Effective Capacity in Block-Fading Channels}

In this section, we seek to identify the constant arrival rates $R$ that can be supported by the two-hop system according to the statistical delay tradeoff characterized earlier. We
consider a block fading scenario in which the fading stays constant
for a block of $T$ seconds and changes independently from one block
to another.

We assume that the channel state information (CSI) of the link $\bS-\bR$ is available at $\bS$ and $\bR$, and the CSI of the link $\bR-\bD$ is available at $\bR$ and $\bD$. The instantaneous capacities of the $\bS-\bR$ and $\bR-\bD$ links in each block are given, respectively, by
\begin{gather}\label{eq:channelcap}
C_1= TB\log_2(1+\tsnr_1
z_1), \quad \text{ and } \quad
C_2=  TB\log_2(1+\tsnr_2
z_2),
\end{gather}
in the units of bits per block or equivalently bits per $T$ seconds. These can be regarded as the service processes at the source and relay.

%

\subsection{Buffer Stability and Log-Moment Generating Function of Block Fading Channels}

To ensure the stability of the queues, we need to enforce the following condition \cite{chang2}
\begin{align}
\E_{z_1}\{C_{1}\} < \E_{z_2}\{C_{2}\},\label{eq:bufferstab2}
\end{align}
i.e., the average arrival rate for the queue at the relay should be less than the average service rate.

Under the block fading assumption, the LMGFs for the service processes of queues at the source $\bS$ and the relay $\bR$ as
functions of $\theta$ are given by 
\begin{align}\label{eq:preequ-1}
\Lambda_{sr}(\theta)&=\log \E_{z_1}\left\{e^{\theta C_{1}}\right\}, \quad\text{and}\quad
\Lambda_{rd}(\theta)=\log \E_{z_2}\left\{e^{\theta C_{2}}\right\}.
\end{align}
The LMGF for the arrival process of the queue at the relay is \cite{deli-twohop}
\begin{align}
\Lambda_{r}(\theta) =\left\{
\begin{array}{ll}
 R\theta,& 0\le\theta\le \tilde{\theta},\\
 R\theta + \log\E_{z_1}\left\{e^{ \left( \theta - \tilde{\theta}\right) C_1}\right\},&\theta> \tilde{\theta}.
\end{array}
\right.
\end{align}


\subsection{Effective Capacity under Statistical Delay Constraints}\label{sec:fixed}

In the following, we first assume that there exist $\theta_1$ and $\theta_2$ such that (\ref{eq:queue12cond}) is satisfied. We can identify the effective capacity associated with the given $\theta_1$ and $\theta_2$ values from Theorem \ref{theo:fixed}. Reminding the statistical delay tradeoff indicated in Lemma \ref{lemm:J1J2relation}, we can obtain the maximum effective capacity by looping over all possible $(J_1(\theta_1),J_2(\theta_2))$, i.e., $\theta_1$ and $\theta_2$, which is the effective capacity under the statistical delay constraint in Definition \ref{def:ecdef}.

From (\ref{eq:J1J2eq}) and (\ref{eq:preequ-1}), we have
\begin{align}\label{eq:J1}
J_1(\theta) =  -\log\E_{z_1}\{e^{-\theta C_1}\},\,\,\text{and}\,\,J_2(\theta) =  -\log\E_{z_2}\{e^{-\theta C_2}\}.
\end{align}
We can show the following properties of $J(\theta)$.
\begin{Lemm}\label{lemm:J1}
Consider the function
\begin{align}
J(\theta) = -\log\E_z\{e^{-\theta C}\}\,\quad\text{for}\,\quad\theta\ge0,
\end{align}
where $C=TB\log_2(1+\tsnr z)$.
This function has the following properties.
\begin{enumerate}[a)]

\item $J(0) = 0$.

\item $J(\theta)$ is increasing in $\theta$, and $\dot{J}(0) = \E_z\{C\}>0$, i.e., the first derivative of $J(\theta)$ with respect to $\theta$ at $\theta = 0$ is given by the average service rate.

\item $J(\theta)$ is a concave function of $\theta$.

\item $\lim_{\theta\to\infty}J(\theta) = -\log \Pr\{C=0\}$, i.e., the negative of the logarithm of the probability of the event that the service rate is 0.

\end{enumerate}
\end{Lemm}
\emph{Proof: }See Appendix \ref{app:J1}.

\begin{Rem}
From the properties above, we can see that $J(\theta)$ is equal to 0 at $\theta=0$, and then it increases sublinearly, and approaches an upperbound, if it exists, as $\theta\to\infty$. Therefore, $J(\theta)$ is a bijective function of $\theta$, and for each value of $J$, we can find the associated $\theta$. Note that the effective capacity expressed as $\frac{J(\theta)}{\theta}$ is decreasing in $\theta$ \cite{deli-twohop}.
\end{Rem}

\begin{Rem}
In the remainder of the paper, we use the following definitions
\begin{align}
R_1&=\frac{J_1(\theta_1)}{\theta_1},\,\,\quad\text{and}\,\,\quad
R_2=\frac{J_2(\theta_2)}{\theta_2}.\label{eq:Rdef}
\end{align}
\end{Rem}

%
%
%
%
%

\begin{Assu}
Throughout this paper, we consider the fading distributions that satisfy the following conditions: 1) $\Pr\{z_{1}=0 \}=0$; 2) $\Pr\{z_{2}=0\}=0$.
\end{Assu}

\begin{Rem}
Under the above assumption, we can see that $J_1(\theta)$ and $J_2(\theta)$ approaches $\infty$ as $\theta$ increases. Note that for the continuous distributions of the fading states, such as Rayleigh and Rician fading, the above assumption is justified immediately. If the above assumption does not hold, we can see that the upper bounds for $J_1(\theta_1)$ and $J_2(\theta_2)$ are finite-valued, and the following analysis still holds while only considering a sliced part of $(J_1,J_2)$ of the $J_1-J_2$ curve characterized in Lemma \ref{lemm:J1J2relation}.
\end{Rem}


\begin{Rem}
According to Lemma \ref{lemm:J1} and the conditions specified in (\ref{eq:cond1}) and (\ref{eq:cond2}), we can see that the effective capacity obtained always satisfies the statistical delay constraints as long as $\theta_1$ and $\theta_2$ satisfy (\ref{eq:queue12cond}). Therefore, with the definitions of $J_1(\theta_1)$ and $J_2(\theta_2)$ in (\ref{eq:J1}), we can find the associated $\theta_1$ and $\theta_2$ on the lower boundary curve indicated by Lemma \ref{lemm:J1J2relation}. Iterating over this set of $\theta_1$ and $\theta_2$, we can derive the maximum effective capacity under end-to-end statistical delay constraints. For other values of $\theta_1$ and $\theta_2$, either (\ref{eq:queue12cond}) cannot be satisfied, or one of the queues is subject to a more stringent constraint than necessary, decreasing the achievable throughput.
\end{Rem}

For the following analysis, we define
\begin{align}\label{eq:omegaeps}
\Omega_{\varepsilon} = \{(\theta_1,\theta_2): \text{ $J_1(\theta_1)$ and $J_2(\theta_2)$ are solutions to }\,(\ref{eq:J1J2function})\}.
\end{align}

We can characterize the effective capacity of the two-hop system given the statistical queueing constraints $\theta_1$ and $\theta_2$ in Theorem \ref{theo:fixed}. Now, we are seeking to identify the effective capacity of the two-hop system under statistical delay constraints specified by $(\varepsilon,D_\tmax)$, in which case $\theta_1$ and $\theta_2$ are unknown. Combining the behavior of $R_E(\theta_1,\theta_2)$ given $(\theta_1,\theta_2)$ and the tradeoff between $J_1(\theta_1)$ and $J_2(\theta_2)$ in Lemma \ref{lemm:J1J2relation}, we have the following result. Note that $z_{i,\tmin}$ and $z_{i,\tmax}$ denote the minimum and maximum value of $z_i$, respectively.
\begin{Lem}\label{theo:ecresultfix}
The effective capacity of the two-hop wireless communication system subject to end-to-end statistical delay constraints specified by $(\varepsilon,D_{\tmax})$ is given by the following:

\vspace{.3cm}
\textbf{\underline{Case I}}: If $\theta_{1,th}= \theta_{2,th}$,
\begin{gather}
\hspace{-.5cm}R_\varepsilon(\varepsilon,D_\tmax)=\frac{J_{th}(\varepsilon)}{\theta_{1,th}},
\end{gather}
where ($\theta_{1,th}$,$\theta_{2,th}$) is the unique solution pair to $J_1(\theta_1)=J_{th}(\varepsilon)$, and $J_2(\theta_2)=J_{th}(\varepsilon)$.

\textbf{\underline{Case II}}: If $\theta_{1,th}>\theta_{2,th}$,
\begin{gather}\label{eq:ecresultcase2}
\hspace{-.5cm}R_\varepsilon(\varepsilon,D_\tmax)= \left\{
\begin{array}{ll}
\frac{J_0}{\theta_{1,0}},&TB\log_2(1+\tsnr_2 z_{2,\tmin})\ge TB\log_2(1+\tsnr_1 z_{1,\tmax}),\\
\frac{J_{1}(\vvtheta_1)}{\vvtheta_{1}},&\text{otherwise.}
\end{array}
\right.
\end{gather}
where $\theta_{1,0}$ is the solution to $J_1(\theta_1) = J_0$,
and $\vvtheta_1$ is the smallest value of $\theta_1$ with $(\theta_1,\theta_2)\in\Omega_\varepsilon$ satisfying
\begin{align}\label{eq:fixresultcond}
&-\frac{1}{\theta_1}\log\E_{z_1}\left\{e^{-\theta_1 C_1}\right\}
=-\frac{1}{\theta_1}\Big(\log\E_{z_2}\left\{e^{-\theta_2
C_{2}}\right\}
+\log\E_{z_1}\left\{e^{(\theta_2-\theta_1)
C_1}\right\}\Big).
\end{align}
Moreover, if $\frac{d J_2(\theta)}{d\theta}\big|_{\theta=\bbtheta_1}\le \frac{d J_1(\theta)}{d\theta}\big|_{\theta=\bbtheta_1}$, where $\bbtheta_1$ is the value of $\theta_1$ with $(\theta_1,\theta_2)\in\Omega_\varepsilon$ satisfying
\begin{align}
\theta_1=\theta_2,
\end{align}
the solution to (\ref{eq:fixresultcond}) with $(\theta_1,\theta_2)\in\Omega_\varepsilon$ is unique.

\textbf{\underline{Case III}}: If $\theta_{1,th}<\theta_{2,th}$,
\begin{gather}\label{eq:ecresultcase3}
\hspace{-.5cm}R_\varepsilon(\varepsilon,D_\tmax)=
\left\{
\begin{array}{ll}
\frac{J_0}{\theta_{2,0}},&TB\log_2(1+\tsnr_1 z_{1,\tmin})\ge\frac{J_0}{\theta_{2,0}}\\
\frac{J_{2}(\uutheta_2)}{\uutheta_2},&\text{otherwise.}
\end{array}
\right.
\end{gather}
where $\theta_{2,0}$ is the solution to $J_2(\theta_2) = J_0$, and ($\uutheta_1$,$\uutheta_2$) is the unique solution to
\begin{align}
\frac{J_{1}(\theta_1)}{\theta_1} = \frac{J_{2}(\theta_2)}{\theta_2}
\end{align}
with $(\theta_1,\theta_2)\in\Omega_\varepsilon$.
\end{Lem}
\emph{Proof:} See Appendix \ref{app:ecresultfix}.

\begin{Rem}
The above theorem covers all the possibilities in which symmetric or asymmetric delay constraints on the queues at the source and relay nodes can be optimal in the sense of achieving the maximum effective capacity of the two-hop relay system. \textbf{Case I} refers to the case that the maximum throughput can be achieved with symmetric delay constraints at the queues of the source and relay. \textbf{Case II} represents the case when the statistical delay constraints at the relay can be more stringent, while \textbf{Case III} shows the scenario with stricter delay constraints at the source. Recalling Theorem \ref{theo:fixed}, we know that as $\varepsilon\to1$, $\theta_1\to0$ and $\theta_2\to0$, and hence
\begin{align}
\lim_{\varepsilon\to1}R_\varepsilon(\varepsilon,D_\tmax)&=\min\left\{\lim_{\theta_1\to0}\frac{J_1(\theta_1)}{\theta_1},\lim_{\theta_1\to0}\frac{J_2(\theta_2)}{\theta_2}\right\}\\
&=\min\left\{\E\{C_1\},\E\{C_2\}\right\}.
\end{align}
\end{Rem}

\section{Numerical Results}

\begin{figure}
\begin{center}
\includegraphics[width=\figsize\textwidth]{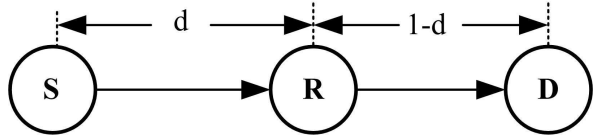}
\caption{The relay model.}\label{fig:systemmodel2}
\end{center}
\end{figure}

\begin{figure}
\begin{center}
\subfigure[Effective capacity vs. $\tsnr_2$. $\tsnr_1=0$ dB. $\varepsilon=0.05$.]{\label{fig:ecinsnr_eps1e-3_dmax=1}
\includegraphics[width=\subfigsize\textwidth]{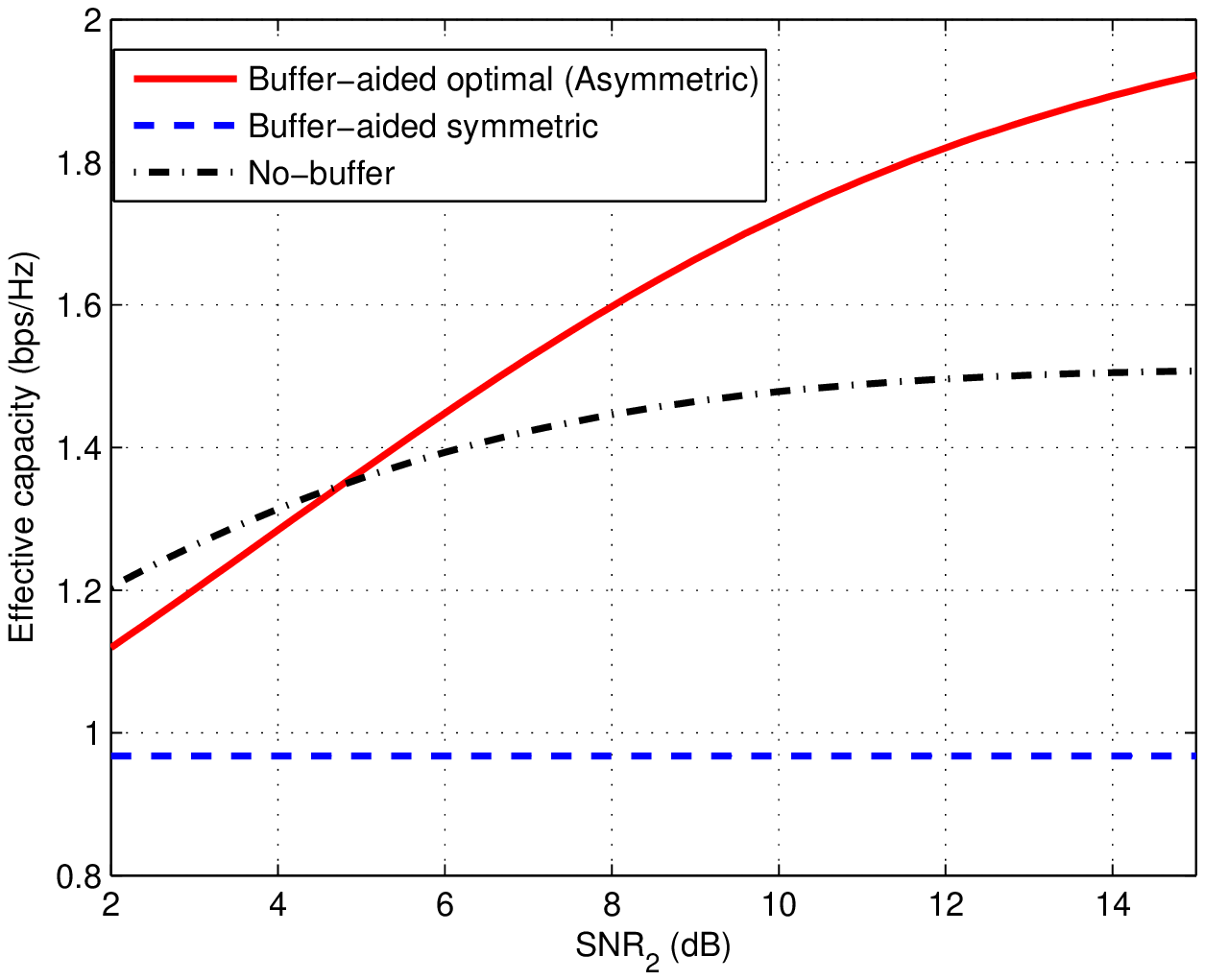}}
\subfigure[$J_2(\theta_2)$ vs. $J_1(\theta_1)$ as $\tsnr_2$ varies. $\tsnr_1=0$ dB. $\varepsilon=0.05$.]
{\label{fig:J1J2change_SNR_eps1e-3_dmax=1}
\includegraphics[width=\subfigsize\textwidth]{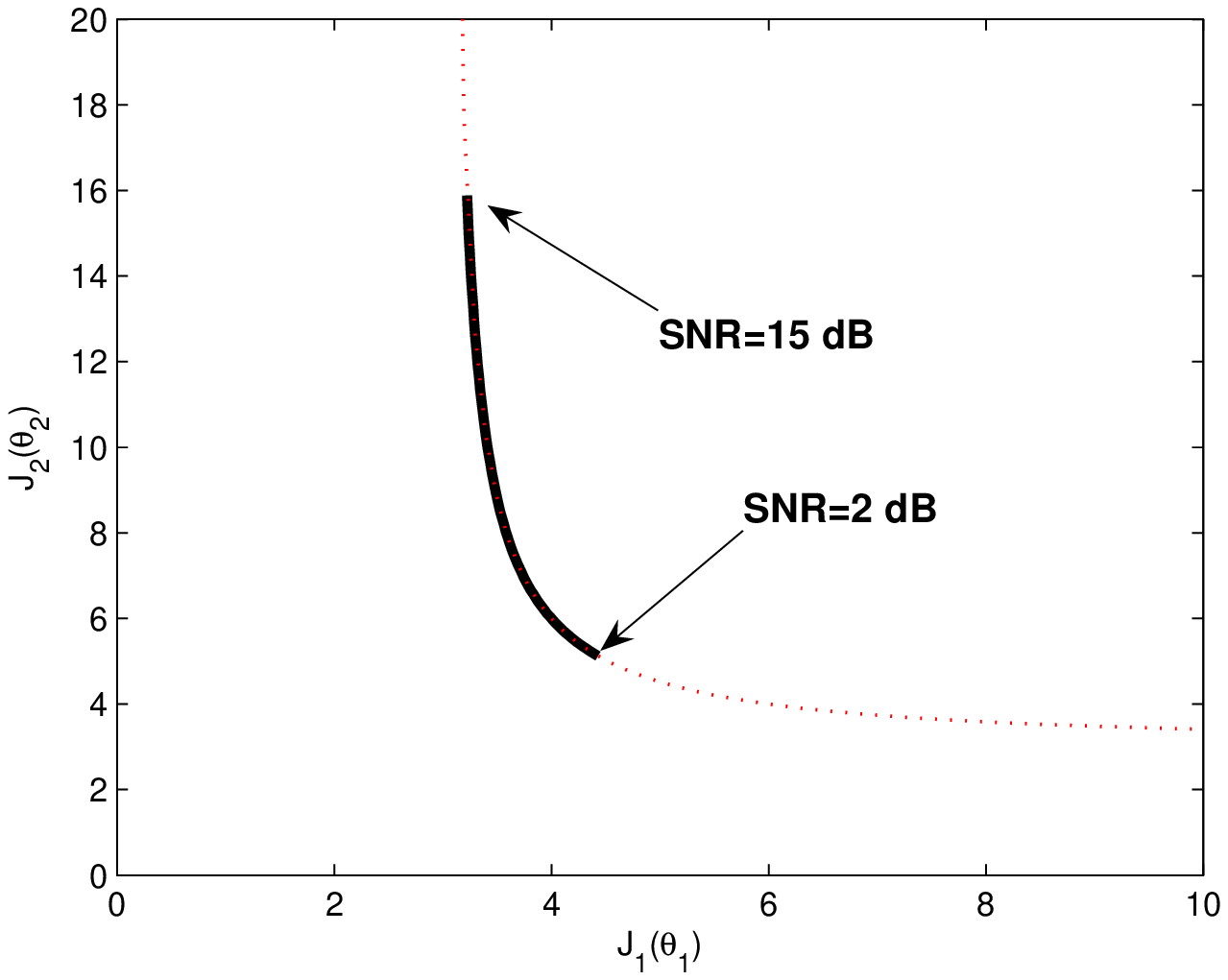}
}
\caption{Effective capacity as a function of $\tsnr_2$.}
\end{center}
\end{figure}

We consider the relay model depicted in Fig. \ref{fig:systemmodel2}.
The source, relay, and destination nodes are located on a straight
line. The distance between the source and the destination is
normalized to 1. Let the distance between the source and the relay
node be $d \in(0,1)$. Then, the distance between the relay and
the destination is $1-d$. We assume the fading distributions for
$\bS-\bR$ and $\bR-\bD$ links follow independent Rayleigh fading with means
$\E\{z_1\} = 1/d^\alpha$ and $\E\{z_2\} = 1/(1-d)^\alpha$, respectively, where we assume that the path loss $\alpha=4$.
We assume that $D_\tmax=1$ sec, and $\tsnr_1=0$ dB in the following numerical
results. The curve ``Buffer-aided optimal (Asymmetric)'' stands for the results in Theorem \ref{theo:ecresultfix}. We also plot the achievable rate when there is no buffer at the relay node ``No-buffer'' \cite{tangrelay}, i.e., the service rate of the queue at the source is given by $\frac{TB}{2}\min\{\log_2(1+2\tsnr_1 z_1), \log_2(1+2\tsnr_2 z_2)\}$ \cite{coopdiver}, and the effective capacity with symmetric delay constraints for the two queues ``Buffer-aided symmetric'', i.e., $J_1(\theta_1)=J_2(\theta_2) = J_{th}(\varepsilon)$ \cite{khalekrelay}, \cite{durelay}.

In Fig. \ref{fig:ecinsnr_eps1e-3_dmax=1}, we plot the effective capacity as a
function of \tsnr\, of the relay node. We fix $d=0.5$, in which case the  $\bS-\bR$ and $\bR-\bD$ links experience the same channel
conditions on average. We assume that the maximum delay violation probability is $\varepsilon=0.05$. From the figure, we can see that the effective capacity of the two-hop system increases with $\tsnr_2$. Note that at small values of $\tsnr_2$, the buffer at the relay introduces certain loss in the achievable rate. As $\tsnr_2$ increases, the buffer at the relay can be beneficial to the two-hop system under statistical delay constraints such that the achievable throughput can be larger. And, in all cases, the achievable rate of asymmetric delay constraints is greater than the one achieved with symmetric delay constraints at the two buffers. In Fig. \ref{fig:J1J2change_SNR_eps1e-3_dmax=1}, we plot the associated $J_2(\theta_2)$ as a function of $J_1(\theta_1)$. As can be seen from the figure, $J_2(\theta_2)$ increases as $\tsnr_2$ increases, i.e., we can impose more stringent constraints to the queue at the relay, and hence the delay constraint at the source can be relaxed. In this way, the effective capacity of the two-hop system can be improved.

\begin{figure}
\begin{center}
\subfigure[Effective capacity v.s. $\varepsilon$. $\tsnr_1=0$ dB.]
{\label{fig:ecindelaybound}
\includegraphics[width=\subfigsize\textwidth]{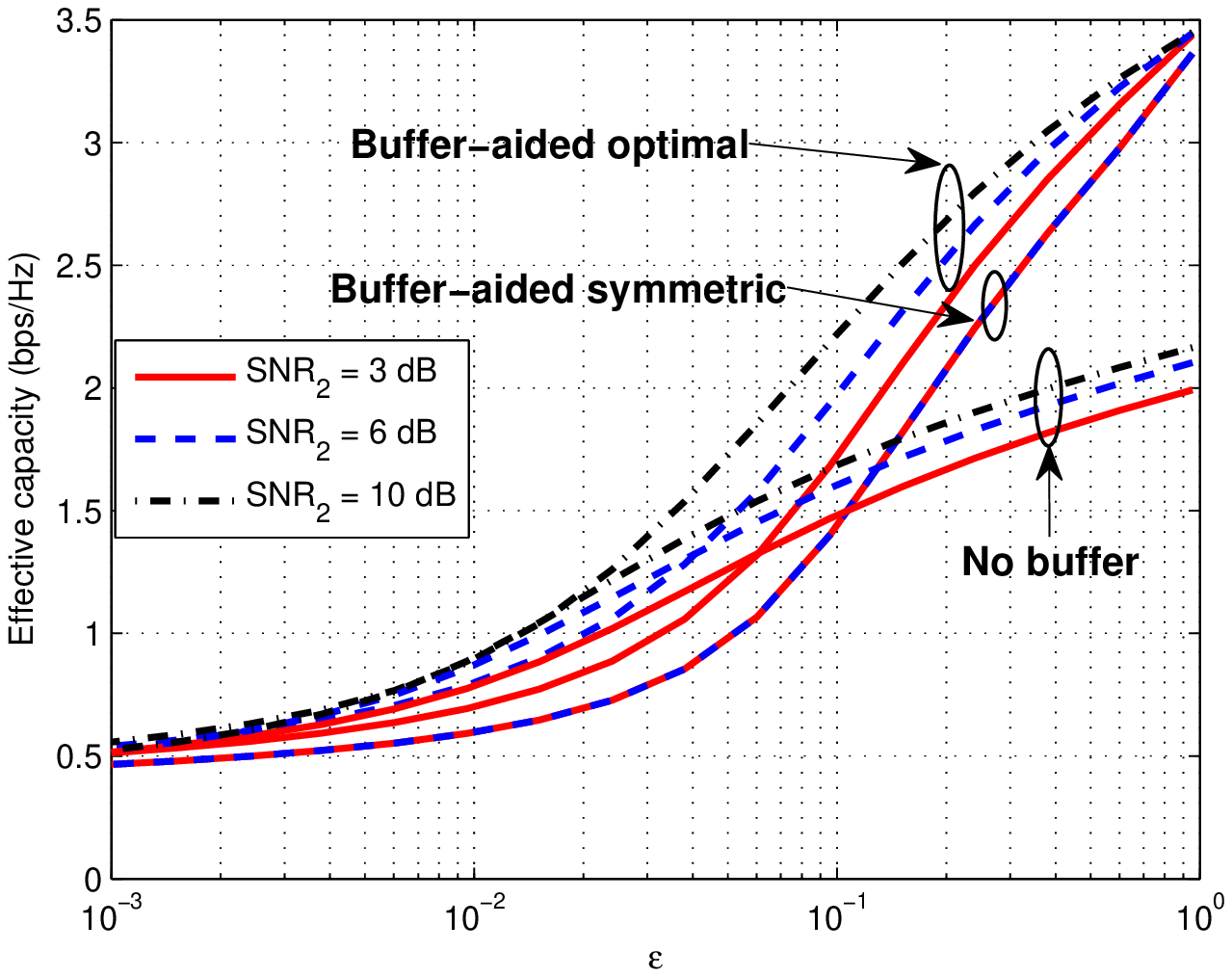}
}
\subfigure[$J_2(\theta_2)$ v.s. $J_1(\theta_1)$ as $\varepsilon$ varies. $\tsnr_1=0$ dB.]{
\label{fig:J1J2change_delaybound}
\includegraphics[width=\subfigsize\textwidth]{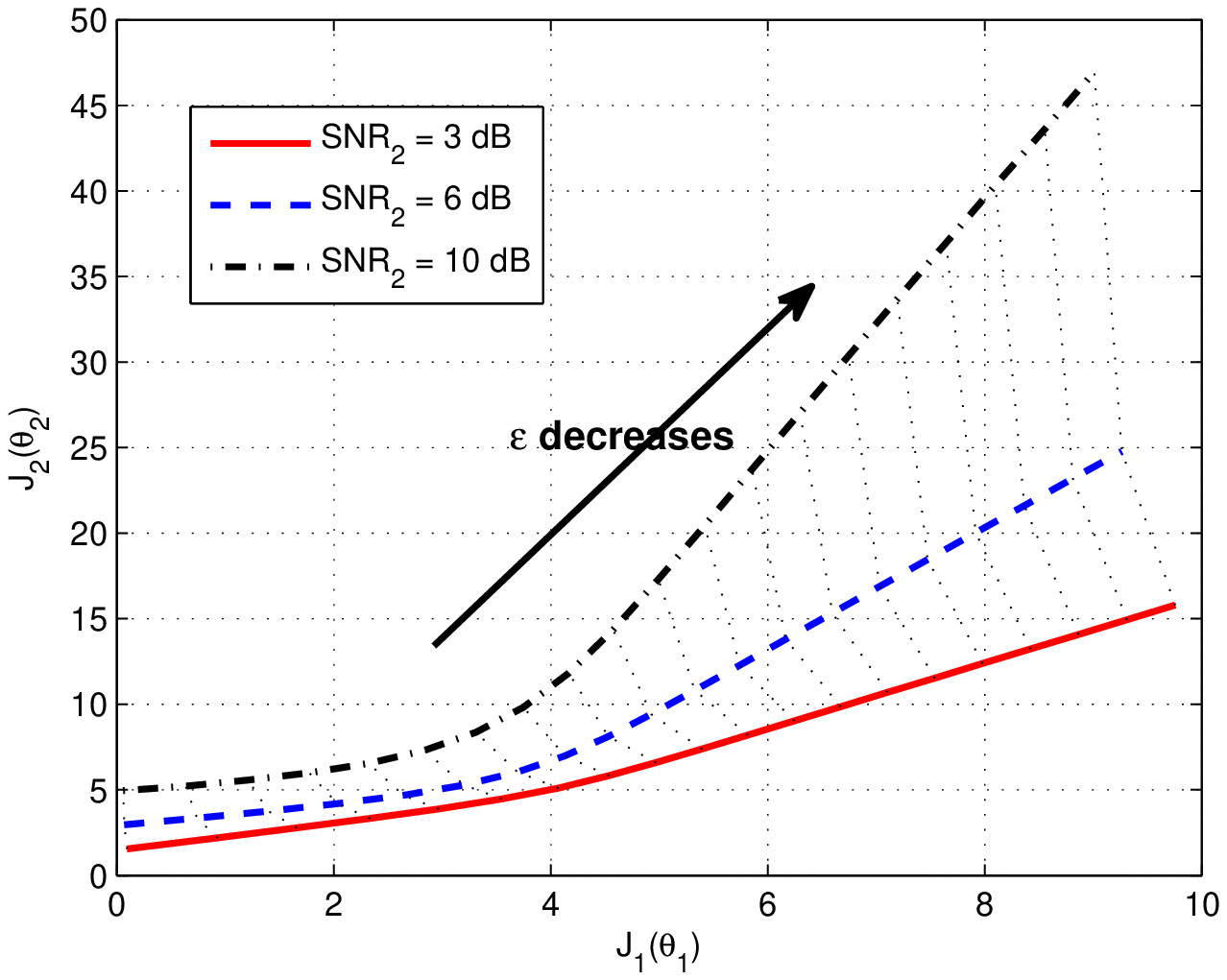}}
\caption{Effective capacity as a function of $\varepsilon$.}
\end{center}
\end{figure}

We are also interested in the impact of the delay violation probability $\varepsilon$ on the achievable performance. In Fig. \ref{fig:ecindelaybound}, we plot the effective capacity as $\varepsilon$ varies for $\tsnr_2 = \{3, 6, 10\}$ dB. It is not surprising that when $\varepsilon\to1$, the effective capacities for different $\tsnr_2$ are the same, since $R_\varepsilon(\varepsilon,D_\tmax)\to\min\{\E\{C_1\},\E\{C_2\}\}=\E\{C_1\}$ in this case. Also, when $\varepsilon\to1$, the achievable rate with buffer at the relay is larger than the achievable rate without buffer at the relay, in accordance with the finding in \cite{bufferrelay} that the throughput can be improved by buffer-aided relay. Moreover, it is interesting that when $\varepsilon$ is relatively large but not one, i.e., the statistical delay constraints are less stringent, the achievable throughput with buffer at the relay is larger. Therefore, buffer-aided relay can be helpful even in the presence of end-to-end delay constraints for certain cases. Also, we can find that for larger $\tsnr_2$, the buffer at the relay can help improve the achievable rate at a smaller $\varepsilon$, i.e., in the presence of more stringent delay constraints. To get more insights, we also plot the associated values of $J_1(\theta_1)$ and $J_2(\theta_2)$ as $\varepsilon$ decreases in Fig. \ref{fig:J1J2change_delaybound}. We can see that the increase in $J_2(\theta_2)$ becomes larger in comparison with $J_1(\theta_1)$. Considering the convexity of $J_2(\theta_2)$ in $J_1(\theta_1)$ in Lemma \ref{lemm:J1J2relation}, loosening the queueing constraint at one queue will require the other queue to operate in a much more conservative way, which provides little gain under more stringent delay constraints, i.e., for smaller $\varepsilon$.
\begin{figure}
\begin{center}
\includegraphics[width=\subfigsize\textwidth]{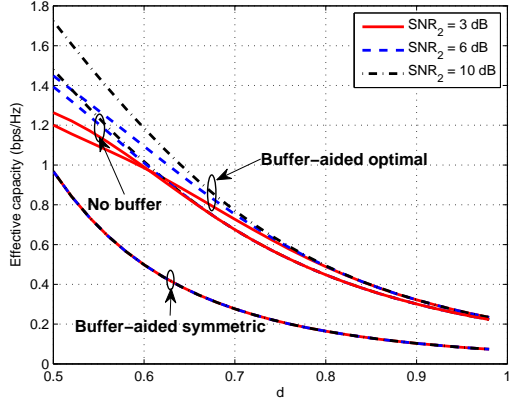}
\caption{Effective capacity as a function of $d$. $\tsnr_1=0$ dB. $\varepsilon=0.05$.}\label{fig:ecind_epsilon=001}
\end{center}
\end{figure}

\begin{figure}
\begin{center}

%
\subfigure[Effective capacity v.s. $d$ and $\varepsilon$. $\tsnr_1=0$ dB. $\tsnr_2=3$ dB.]
{\label{fig:ecindelayboundd}\includegraphics[width=\subfigsize\textwidth]{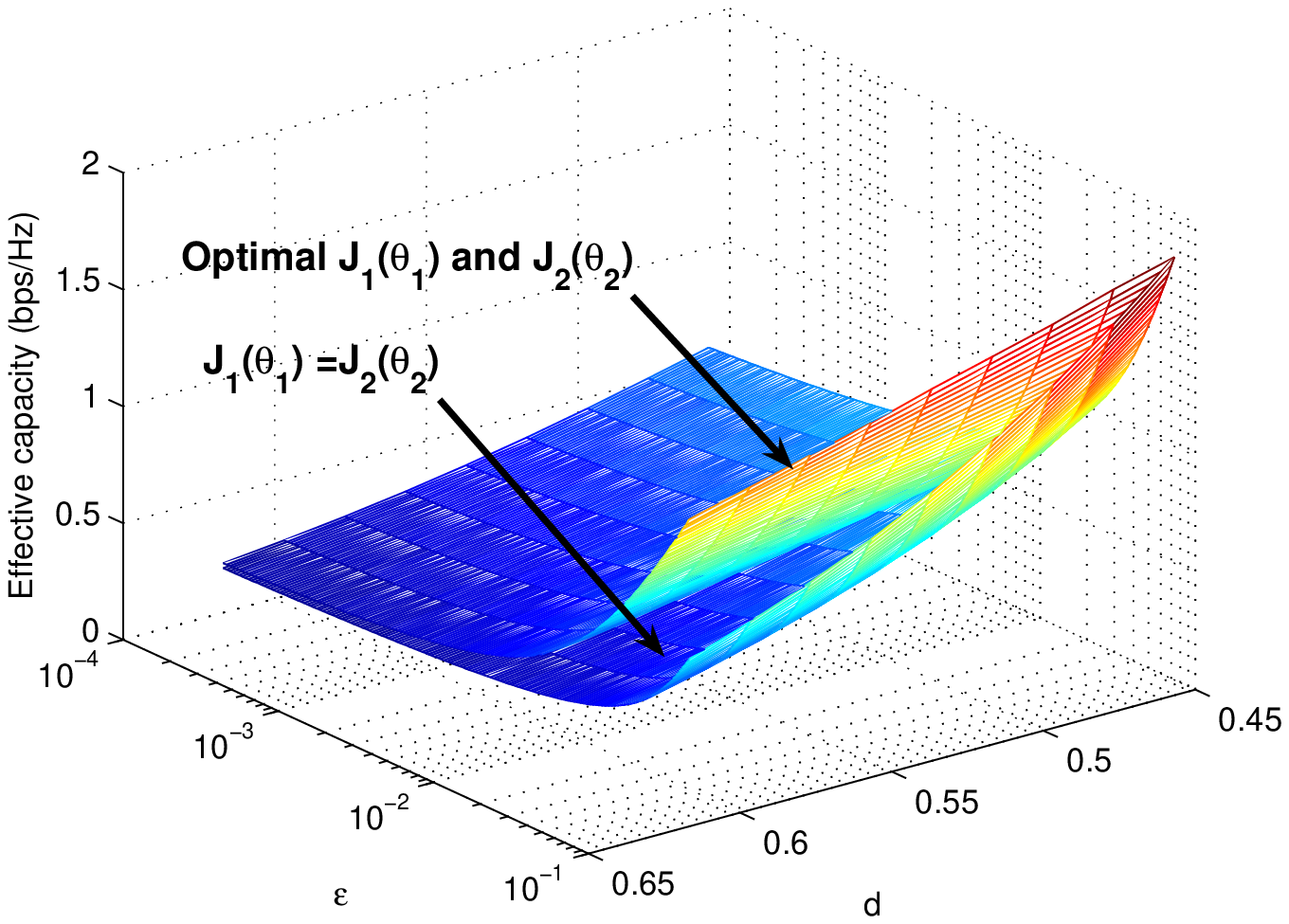}}
\subfigure[$J_1(\theta_1)$ and $J_2(\theta_2)$ as functions of $d$ and $\varepsilon$. $\tsnr_1=0$ dB. $\tsnr_2=3$ dB. ]
{\label{fig:J1J2indelayboundd}\includegraphics[width=\figsize\textwidth]{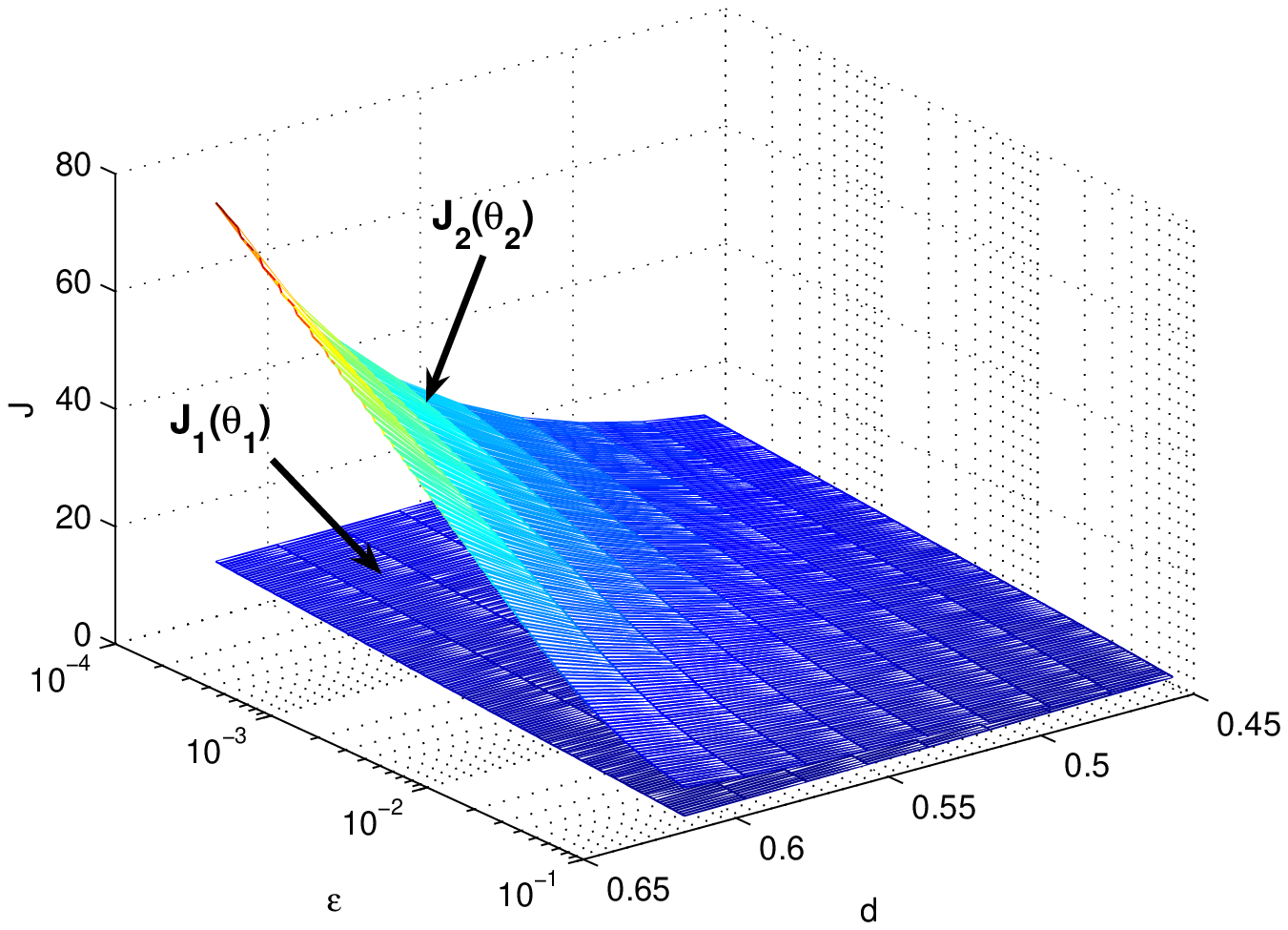}}
\caption{Effective capacity as a function of $d$ and $\varepsilon$.}
\end{center}
\end{figure}

 In Fig. \ref{fig:ecind_epsilon=001}, we plot the effective capacity as $d$ varies.  We assume $\tsnr_2=\{3,6,10\}$ dB, $\varepsilon=0.05$. We can see from the figure that as $d$ increases, i.e., the channel condition at the link $\bS-\bR$ is worse, the effective capacity decreases, and the increase of \tsnr\, at the relay node helps little. It is interesting that even for small values of $\tsnr_2$, as $d$ increases, the buffer at the relay can help improve the achievable throughput. Albeit, the benefits provided by the buffer at the relay vanish as $d$ approaches 1 since the link $\bS-\bR$ becomes the bottleneck of the system. Finally, we plot the effective capacity as $d$ and $\varepsilon$ vary in Fig. \ref{fig:ecindelayboundd}, with the associated delay tradeoff $J_1(\theta_1)$ and $J_2(\theta_2)$ for the proposed asymmetric delay constraints in Fig. \ref{fig:J1J2indelayboundd}. We assume $\tsnr_2 = 3$ dB. As can be seen from the figure, for all cases, effective capacity decreases as $d$ increases or $\varepsilon$ decreases. The improvement in effective capacity is achieved through strong bias towards the queue at the source, in which case we have much larger $J_2(\theta_2)$ in comparison with $J_1(\theta_1)$. 


%
%

\section{Conclusion}
In this paper, we have investigated the maximum constant arrival rates that can
be supported by a two-hop communication link with a buffer-aided relay under end-to-end statistical delay constraints. We have provided a unified framework for achieving the statistical delay tradeoffs imposed to the source and relay nodes while satisfying the statistical delay constraints. We have determined
the effective capacity in the block-fading scenario as a function of the statistical delay constraints,
the signal-to-noise ratio levels $\tsnr_1$ and $\tsnr_2$, and the fading distributions. We have shown that asymmetric delay constraints at the two buffers can help increase the effective capacity of the two-hop system compared with symmetric delay constraints. We have found that buffer-aided relay can improve the achievable rate of the system under delay constraints when the $\tsnr$ at the relay is high, the end-to-end delay constraints is loose, or when the channel conditions between the relay and destination node are more favorable.

\vspace{-.5cm}
\appendix
\subsection{Preliminary Results}\label{app:prev}

\begin{Lem1}(\cite{deli-twohop})\label{prop:upperbound}
The constant arrival rates, which can be supported by the two-hop link in the presence of queueing constraints $\theta_1$ and $\theta_2$ at the source and relay, respectively, are upperbounded by
\begin{align}\label{eq:upperboundrate}
R\le\min\left\{-\frac{1}{\theta_1}\log\E_{z_1}\left\{e^{-\theta_1C_1}\right\}, -\frac{1}{\theta_2}\log\E_{z_2}\left\{e^{-\theta_2C_2}\right\}\right\}
=\min\left\{\frac{J_1(\theta_1)}{\theta_1}, \frac{J_2(\theta_2)}{\theta_2}\right\}.
\end{align}
\end{Lem1}

\begin{Lem}(\cite{deli-twohop}) \label{theo:fixed}
The effective capacity of the two-hop system given $\theta_1>0$ and $\theta_2>0$ is given by the following:
\vspace{.3cm}

\textbf{\underline{Case I}}: If $\theta_1\ge \theta_2$,
\begin{gather}
\hspace{-.5cm}R_E(\theta_1,\theta_2)=\min\Bigg\{-\frac{1}{\theta_1}\log\E_{z_1}\left\{e^{-\theta_1
C_1}\right\},
-\frac{1}{\theta_2}\log \E_{z_2}\left\{e^{-\theta_2
C_{2}}\right\}\Bigg\}.
\end{gather}

\textbf{\underline{Case II}}: If $\theta_1<\theta_2$ and $\theta_2\le \btheta$,
\begin{gather}
R_E(\theta_1,\theta_2)=-\frac{1}{\theta_1}\log\E_{z_1}\left\{e^{-\theta_1
C_1}\right\} \label{eq:theorem1case2part1}
\end{gather}
where $\btheta$ is the unique value of $\theta$ for which we have the following equality satisfied:
\begin{align}\label{eq:fixcase2cond}
&-\frac{1}{\theta_1}\log\E_{z_1}\left\{e^{-\theta_1 C_1}\right\}
=-\frac{1}{\theta_1}\Big(\log\E_{z_2}\left\{e^{-\theta
C_2}\right\}
+\log\E_{z_1}\left\{e^{(\theta-\theta_1)
C_1}\right\}\Big).
\end{align}

\textbf{\underline{Case III}}: Assume $\theta_1<\theta_2$ and $\theta_2 > \btheta$.
\begin{description}
\item{\textbf{III.a}:} If
\begin{align} \label{eq:fixcase3aprecond}
-\frac{1}{\theta_2}\log\E_{z_2}\left\{e^{-\theta_2
C_{2}}\right\} \ge -\frac{1}{\theta_2}\log\E_{z_1}\left\{e^{-\theta_2
C_1}\right\},
\end{align}
then
\begin{gather}
R_E(\theta_1,\theta_2)= -\frac{1}{\tilde{\theta}^*}\log\E_{z_1}\left\{e^{-\tilde{\theta}^*
C_1}\right\}
\end{gather}
where $\tilde{\theta}^*$ is the smallest solution to
\begin{align}\label{eq:fixcase3acond}
&-\frac{1}{\tilde{\theta}}\log\E_{z_1}\left\{e^{-\tilde{\theta}
C_1}\right\}
=-\frac{1}{\tilde{\theta}}\bigg(\log\E_{z_2}\left\{e^{-\theta_2
C_2}\right\}+\log\E_{z_1}\left\{e^{(\theta_2-\tilde{\theta}) C_1}\right\}\Bigg).
\end{align}

\item{\textbf{III.b}:} Otherwise,
\begin{gather}
R_E(\theta_1,\theta_2)= -\frac{1}{\theta_2}\log\E_{z_2}\left\{e^{-\theta_2
C_2}\right\}.
\end{gather}
\end{description}
\end{Lem}

\subsection{Proof of Lemma \ref{lemm:J1J2relation}}\label{app:J1J2relation}

\begin{enumerate}

\item When $J_1(\theta_1)\neq J_2(\theta_2)$, the continuity is obvious since there is no pole to (\ref{eq:J1J2function}). Consider $J_1(\theta_1)=J_2(\theta_2)$. We can see that
     \begin{small}
     \begin{align}
     \hspace{-.5cm}\lim_{J_2(\theta_2)\to J_1(\theta_1)_{-}}\vartheta(J_1(\theta_2),J_2(\theta_2)) &= \lim_{J_2(\theta_2)\to J_1(\theta_1)_{-}}\frac{J_2(\theta_2)e^{-J_1(\theta_1)D_\tmax} - J_1(\theta_1)e^{-J_2(\theta_2)D_\tmax}}{J_2(\theta_2)-J_1(\theta_1)}\\
     & = \lim_{J_2(\theta_2)\to J_1(\theta_1)_{-}}e^{-J_2(\theta_2)D_\tmax} \frac{J_2(\theta_2)e^{-(J_1(\theta_1)-J_2(\theta_2))D_\tmax} - J_1(\theta_1)}{J_2(\theta_2)-J_1(\theta_1)}\\
     & = \lim_{J_2(\theta_2)\to J_1(\theta_1)_{-}}e^{-J_2(\theta_2)D_\tmax} \left(1 + J_{2}(\theta_2)\frac{1 - e^{-(J_1(\theta_1)-J_2(\theta_2))D_\tmax}}{J_1(\theta_1)-J_2(\theta_2)}\right)\\
     & = e^{-J_2(\theta_2)D_\tmax} \left(1+J_2(\theta_2)D_\tmax\right).
     \end{align}
     \end{small}
     Similarly, we can show that
     \begin{align}
     \lim_{J_2(\theta_2)\to J_1(\theta_1)_{+}}\vartheta(J_1(\theta_2),J_2(\theta_2)) = e^{-J_1(\theta_1)D_\tmax} \left(1+J_1(\theta_1)D_\tmax\right).
     \end{align}
      From (\ref{eq:delayprob}), we can see that at $J_1(\theta_1)=J_2(\theta_2)$, $\vartheta(J_1(\theta_2),J_2(\theta_2))$ is continuous, i.e., $J_2(\theta_2)=\Phi(J_1(\theta_1))$ is continuous, and from (\ref{eq:queue12cond}), we should have
\begin{align}
\left(1+J_1(\theta_1)D_{\tmax}\right)e^{-J_1(\theta_1)D_{\tmax}} \le \varepsilon
\end{align}
which gives us (\ref{eq:Jfunctioncond}) immediately by solving the above equation with equality.

\item Taking the partial derivative of $\vartheta(J_1(\theta_1),J_2(\theta_2))$ in $J_1(\theta_1)$ and noting that the right-hand-side (RHS) of (\ref{eq:J1J2function}) is constant, we have
\begin{small}
\begin{align}
\frac{\partial \vartheta(J_1(\theta_1),J_2(\theta_2))}{\partial J_1(\theta)} &=\frac{1}{(J_2(\theta_2)-J_1(\theta_1))^2} \bigg(\Big(\dot{J}_2(\theta)e^{-J_1(\theta_1)D_\tmax}-J_2(\theta_2)D_\tmax e^{-J_1(\theta_1)D_\tmax} - e^{-J_2(\theta_2)D_\tmax} \nonumber\\
&\hspace{.5cm}+J_1(\theta)\dot{J}_2(\theta_2)D_\tmax e^{-J_2(\theta_2)D_\tmax}\Big)(J_2(\theta_2)-J_1(\theta_1)) -(\dot{J}_2(\theta_2)-1)\nonumber\\
&\hspace{1cm}\times\left(J_2(\theta_2)e^{-J_1(\theta_1)D_\tmax}-J_1(\theta_1)e^{-J_2(\theta_2)D_\tmax}\right)\bigg)=0,
\end{align}
\end{small}
which, after combining the coefficients of $\dot{J}_2(\theta_2)$ and rearrangements, gives us
\begin{small}
\begin{align}
\dot{\Phi}(J_1(\theta_1))=\dot{J}_2(\theta_2)
& = \frac{J_2(\theta_2)}{J_1(\theta_1)}e^{(J_2(\theta_2)-J_1(\theta_1))D_\tmax}\frac{(J_2(\theta_2)-J_1(\theta_1))D_\tmax+e^{-(J_2(\theta_2)-J_1(\theta_1))D_\tmax}-1}{(J_2(\theta_2)-J_1(\theta_1))D_\tmax+1 - e^{(J_2(\theta_2)-J_1(\theta_1))D_\tmax}}
\end{align}
\end{small}
In the following, we will show that $\dot{\Phi}(J_1(\theta_1))< 0$. Denote $x=(J_2(\theta_2)-J_1(\theta_1))D_\tmax$, and define
\begin{align}
\nu(x) = \frac{x+e^{-x}-1}{x+1-e^{x}}.
\end{align}
Then, we can rewrite $\dot{\Phi}(J_1(\theta))$ as
\begin{align}\label{eq:J2der}
\dot{\Phi}(J_1(\theta_1))=\dot{J}_2(\theta_2)
& = \frac{J_2(\theta_2)}{J_1(\theta_1)}e^{(J_2(\theta_2)-J_1(\theta_1))D_\tmax}\nu(x).
\end{align}
Note that $\frac{J_2(\theta_2)}{J_1(\theta_1)}e^{(J_2(\theta_2)-J_1(\theta_1))D_\tmax}$ is positive. Taking the first derivative of $\nu(x)$, we obtain
\begin{align}\label{eq:J1J2proof1}
\dot{\nu}(x)=\frac{4-2\left(e^{x}+e^{-x}\right)+x\left(e^x-e^{-x}\right)}{\left(x+1-e^{x}\right)^2}
\end{align}
We can show that $\dot{\nu}(x)\ge0$. Suppose $x>0$. Considering the numerator of the above equation, we have
\begin{small}
\begin{align}
4-2\left(e^{x}+e^{-x}\right)+x\left(e^x-e^{-x}\right) &=-2\left(e^{\frac{x}{2}}-e^{-\frac{x}{2}}\right)^2+x\left(e^{\frac{x}{2}}-e^{-\frac{x}{2}}\right)\left(e^{\frac{x}{2}}+e^{-\frac{x}{2}}\right)\\
& = \left(e^{\frac{x}{2}}-e^{-\frac{x}{2}}\right)\left(-2\left(e^{\frac{x}{2}}-e^{-\frac{x}{2}}\right)+x\left(e^{\frac{x}{2}}+e^{-\frac{x}{2}}\right)\right)\\
&=e^{-\frac{x}{2}}(x+2)\left(e^{\frac{x}{2}}-e^{-\frac{x}{2}}\right)\left(\frac{x-2}{x+2}e^{x}+1\right)\\
&\ge 0
\end{align}
\end{small}
where $\frac{x-2}{x+2}e^{x}\ge-1$ is incorporated since it is an increasing function of $x$, and its value at $x=0$ is $-1$. Therefore, $\dot{\nu}(x)>0$ for $x>0$, i.e., $\nu(x)$ is increasing for $x>0$. In a similar way, we can show that $\dot{\nu}(x)>0$ for $x<0$. Additionally, we can show $\lim_{x\to0}\dot{\nu}(x) = 0$ by considering the Taylor expansions of $e^x$ and $e^{-x}$ at $x=0$ and noting that the numerator goes to 0 in the order $o(x^4)$ while the denominator goes to 0 in the order of $x^4$. Therefore, $\nu$ is increasing in $x$. Meanwhile,
\begin{align}
\lim_{x\to\infty}\nu(x) = \lim_{x\to\infty}\frac{x+e^{-x}-1}{x+1-e^{x}}=\lim_{x\to\infty}\frac{1-e^{-x}}{1-e^{x}}=0.
\end{align}
Hence, $\nu(x)<0$, which in turn, tells us that $\dot{\Phi}(J_1(\theta_1)) <0$ in (\ref{eq:J2der}). Therefore, $J_2(\theta_2)=\Phi(J_1(\theta_1))$ is strictly decreasing in $J_1(\theta)$.

\item We will show the convexity of $\Phi$ by considering the branches for $J_2(\theta_2)>J_1(\theta_1)$ and $J_2(\theta_2)<J_1(\theta_1)$, respectively.

    For $J_1(\theta_1)<J_{th}(\varepsilon)$, we know that $J_2(\theta_2)>J_1(\theta_1)$. Consider
\begin{small}
\begin{align}
\dot{J}_2(\theta_2)
& = \frac{J_2(\theta_2)}{J_1(\theta_1)}e^{(J_2(\theta_2)-J_1(\theta_1))D_\tmax}\frac{(J_2(\theta_2)-J_1(\theta_1))D_\tmax+e^{-(J_2(\theta_2)-J_1(\theta_1))D_\tmax}-1}{(J_2(\theta_2)-J_1(\theta_1))D_\tmax+1 - e^{(J_2(\theta_2)-J_1(\theta_1))D_\tmax}}\\
& =\frac{J_2(\theta_2)}{J_1(\theta_1)}e^{x}\nu(x)
\end{align}
\end{small}
where again $x=(J_2(\theta_2)-J_1(\theta_1))D_\tmax$. Note that as $x$ increases, $\frac{J_2(\theta_2)}{J_1(\theta_1)}$ should increase since $J_1(\theta_1)$ decreases and $J_2(\theta_2)$ increases. From the above discussion, we know $\nu(x)<0$, for $x>0$. Define $\eta(x)=e^{x}\nu(x)$, $\eta(x)<0$ for $x>0$. Then, if we can show that $\eta(x)$ is decreasing as $x$ increases, then $\dot{J}_2(\theta_2)=\dot{\Phi}(J_1(\theta_1)) $ will decrease with $x$, since a smaller negative value multiplied with a larger positive value will lead to a smaller negative value. Taking the first derivative of $\eta(x)$, we have
\begin{align}
\dot{\eta}(x) = e^{x}(\nu(x)+\dot{\nu}(x))=e^{x}\frac{2+x^2-(e^{x}+e^{-x})}{\left(x+1-e^{x}\right)^2}.
\end{align}
Note that the numerator $2+x^2-(e^x+e^{-x})$ can be shown to be less than 0 for $x>0$. More specifically, consider that its second derivative $2-(e^{x}+e^{-x})$ is less than 0 for $x>0$ and the first derivative $2x-(e^{x}-e^{-x})$ at $x=0$ is 0, and hence its first derivative is always less than 0, which tells us that it is a decreasing function in $x$ with the maximum value at $x=0$ as 0. Therefore, $\dot{\eta}<0$. Hence, $\dot{J}_2(\theta_2)<0$ is decreasing as $J_1(\theta_1)$ decreases for $J_1(\theta_1)<J_{th}(\varepsilon)$, i.e., $\ddot{\Phi}(J_1(\theta_))\ge 0$. Similarly, we can show that $\ddot{\Phi}(J_1(\theta_1))\ge 0$ for $J_1(\theta_1)>J_{th}(\varepsilon)$. Together, we know that $\ddot{\Phi}(J_1(\theta_1))\ge0$, and hence $J_2(\theta_2)=\Phi(J_1(\theta_1))$ is a convex function in $J_1(\theta_1)$.

\item Letting $J_1(\theta)$ go to infinity, we can see that
\begin{align}
\lim_{J_1(\theta)\to\infty}\vartheta(J_1(\theta),J_2(\theta)) = \lim_{J_1(\theta)\to\infty}e^{-J_2(\theta_2)D_\tmax} = e^{-J_0D_\tmax}
\end{align}
which indicates $\lim_{J_1(\theta)\to\infty}J_2(\theta_2) = J_0$. On the other hand, if we let $J_2(\theta)$ go to infinity, we can show that $\lim_{J_2(\theta)\to\infty}J_1(\theta_1) = J_0$. Together, we obtain the result in the lemma.\hfill$\square$

\end{enumerate}

\subsection{Proof of Lemma \ref{lemm:J1}}\label{app:J1}
\begin{enumerate}[a)]

\item This property can be readily seen by evaluating the function at $\theta = 0$.

\item The first derivative of $J$ with respect to $\theta$ can be evaluated as
\begin{align}
\dot{J}(\theta)=\frac{\E_z\left\{e^{-\theta C}C\right\}}{\E_z\left\{e^{-\theta C}\right\}}>0.
\end{align}
Then, $\dot{J}(0)$ can be obtained by evaluating the above equation at $\theta = 0$.

\item The second derivative of $J$ with respect to $\theta$ can be expressed as
\begin{align}
\ddot{J}(\theta)&=-\frac{1}{\left(\E_{z}\left\{e^{-\theta C}\right\}\right)^2}
\Bigg(\E_z\left\{e^{-\theta C}C^2\right\}\E_z\left\{e^{-\theta C}\right\} 
- \left(\E_z\left\{e^{-\theta C}C\right\}\right)^2\Bigg).
\end{align}

By the Cauchy-Schwarz inequality, we know that
$\E\{X^2\}\E\{Y^2\}\ge\left(\E\{XY\}\right)^2$. Then, denoting
\\
$X=\sqrt{e^{-\theta C} C^2}$ and
$Y=\sqrt{e^{-\theta C}}$, we easily see that
$\ddot{J}(\theta)\le0$ for all $\theta$. Thus, $J(\theta)$ is a concave function.

\item Note that as long as $C\neq 0$, $\lim_{\theta\to\infty}e^{-\theta C}=0$, and whenever $C=0$, $e^{\theta C}=1$. Therefore, we have
    $\lim_{\theta\to\infty}\E_{z\neq0}\left\{e^{-\theta C}\right\} = 0.$
    Then $\lim_{\theta\to\infty}J(\theta) =\lim_{\theta\to\infty}-\log\left(\E_{z\neq 0}\{e^{-\theta C}\}+\E_{z=0}\{1\}\right) = -\log\Pr\{C=0\}$.
\hfill$\square$
\end{enumerate}

\subsection{Proof of Theorem \ref{theo:ecresultfix}}\label{app:ecresultfix}

With the delay tradeoff specified in Lemma \ref{lemm:J1J2relation}, we can see that there is potential improvement of effective capacity by adjusting the statistical delay constraint imposed on the queues at the source and relay nodes. As a starting point, we consider $J_1(\theta_1)=J_2(\theta_2)$. According to Lemma \ref{lemm:J1} and the subsequent discussions, we can always find $\theta_{1,th}$ and $\theta_{2,th}$ for $J_{th}(\varepsilon)$ defined in (\ref{eq:Jfunctioncond}). Now, depending on the values of $\theta_{1,th}$ and $\theta_{2,th}$, we have different behaviors of the effective capacity in Theorem \ref{theo:fixed} in Appendix \ref{app:prev}. We seek to find the optimal $J_1(\theta_1)$ and $J_2(\theta_2)$ with $(\theta_1,\theta_2)\in\Omega_\varepsilon$ to maximize the effective capacity, where $\Omega_\varepsilon$ is defined in (\ref{eq:omegaeps}).

\underline{\textbf{Case I}:} Assume $\theta_{1,th} = \theta_{2,th}$. For this case, we should have
\begin{align}
R_E(\theta_{1,th},\theta_{2,th}) = R_1 = \frac{J_{th}(\varepsilon)}{\theta_{1,th}} = \frac{J_{th}(\varepsilon)}{\theta_{2,th}}=R_2.
\end{align}
We assert that this value is the effective capacity of the two-hop system, i.e.,
\begin{align}
R_\varepsilon(\varepsilon,D_\tmax)=\sup_{(\theta_1,\theta_2)\in\Omega}R_E(\theta_1,\theta_2)=R_E(\theta_{1,th},\theta_{2,th}).
\end{align}
We can show this by contradiction. We know that the effective capacity is a decreasing function in $\theta$. Suppose that there exists some $R>R_E(\theta_{1,th},\theta_{2,th})$ that can be supported by the two-hop system with $\theta_1$ and $\theta_2$. Then, we must have $\theta_1<\theta_{1,th}$, and hence $J_1(\theta_1)<J_1(\theta_{1,th})$. According to the statistical delay tradeoff shown in Lemma \ref{lemm:J1J2relation}, we can see that $J_2(\theta_2)>J_2(\theta_{2,th})$, which tells us that $\theta_2>\theta_{2,th}$ according to Lemma \ref{lemm:J1}, since $J_2(\theta)$ is increasing in $\theta$. Now, from the Proposition \ref{prop:upperbound} in Appendix \ref{app:prev}, we obtain
\begin{align}
R\le\min\left\{\frac{J_1(\theta_1)}{\theta_1},\frac{J_2(\theta_2)}{\theta_2}\right\}=\frac{J_2(\theta_2)}{\theta_2}<\frac{J_2(\theta_{2,th})}{\theta_{2,th}}=R_E(\theta_{1,th},\theta_{2,th})
\end{align}
which leads to a contradiction.

\underline{\textbf{Case II}:} Assume $\theta_{1,th}> \theta_{2,th}$. In this case, we can see that
\begin{align}
R_1 = \frac{J_{1}(\theta_{1,th})}{\theta_{1,th}}=\frac{J_{th}(\varepsilon)}{\theta_{1,th}} <\frac{J_{th}(\varepsilon)}{\theta_{2,th}}= \frac{J_{2}(\theta_{2,th})}{\theta_{2,th}}=R_2.
\end{align}
The effective capacity associated with $\theta_{1,th},\theta_{2,th}$ specializes into \textbf{Case I} of Theorem \ref{theo:fixed}. Therefore, $R_E(\theta_{1,th},\theta_{2,th})=\min\{R_1,R_2\} = R_1$. Obviously, the queueing constraint imposed at the source is more stringent. To achieve better performance, we should try to relax the queueing constraints at the source, i.e., decrease $\theta_1$, or $J_1(\theta_1)$ equivalently. Correspondingly, from Lemma \ref{lemm:J1J2relation}, $J_2(\theta_2)$ should increase, and we have $J_2(\theta_2)>J_{th}(\varepsilon)>J_1(\theta_1)$. In the following, we will provide a characterization of $\theta_1$ as we iterate over $(\theta_1,\theta_2)\in\Omega_\varepsilon$ to find the optimal pair that maximizes the effective capacity.

First, noting that as $J_1(\theta_1)$ decreases from $J_{th}(\varepsilon)$ to $J_0$, we can see that $\theta_1$ decreases from $\theta_{1,th}$ to some finite value $\theta_{1,0}$, which is the solution to $J_1(\theta)=J_0$. To the opposite, $\theta_2$ increases from $\theta_{2,th}<\theta_{1,th}$ to $\infty$. Clearly, from the continuity of $J_2(\theta_2)=\Phi(J_1(\theta_1))$, the corresponding $\theta_2$ as a function of $\theta_1$ should be continuous as well. Hence, there must be one point $(\bbtheta_1,\bbtheta_2)\in\Omega_\varepsilon$ such that
\begin{align}\label{eq:bbthetafix}
\bbtheta_1 = \bbtheta_2,
\end{align}
and for all $(\theta_1,\theta_2)\in\Omega_\varepsilon$ with $\theta_1<\bbtheta_1$, we will have $\theta_2>\bbtheta_2=\bbtheta_1>\theta_1$. According to Lemma \ref{lemm:J1}, we know $J_1(\theta)$ and $J_2(\theta)$ are increasing functions of $\theta$. Therefore, at this point, we have
\begin{small}
\begin{align}
R_1=\frac{J_1(\bbtheta_1)}{\bbtheta_1}<\frac{J_1(\theta_{1,th})}{\bbtheta_1} = \frac{J_{th}(\varepsilon)}{\bbtheta_1}=\frac{J_2(\theta_{2,th})}{\bbtheta_1}<\frac{J_2(\bbtheta_2)}{\bbtheta_1}=\frac{J_2(\bbtheta_2)}{\bbtheta_2}=R_2.
\end{align}
\end{small}
That is, the queue at the source is still the bottleneck of the two-hop system. We can further relieve the queueing constraint at the source.

Now, as $\theta_1$ further decreases, $\theta_1<\theta_2$. Consequently, the effective capacity associated with $(\theta_1,\theta_2)\in\Omega_\varepsilon$ now specializes into \textbf{Case II} of Theorem \ref{theo:fixed}. As can be seen from Theorem \ref{theo:fixed}, the queue at the relay will not affect the performance as long as $\theta_1$ and $\theta_2$ satisfy the following inequality given by
\begin{small}
\begin{align}\label{eq:case2proof1}
&-\frac{1}{\theta_1}\log\E_{z_1}\left\{e^{-\theta_1 C_1}\right\}
\le-\frac{1}{\theta_1}\Big(\log\E_{z_2}\left\{e^{-\theta_2
C_2}\right\}
+\log\E_{z_1}\left\{e^{(\theta_2-\theta_1)
C_1}\right\}\Big).
\end{align}
\end{small}
Note that as $\theta_1$ decreases from $\bbtheta_1$ to $\theta_{1,0}$, the LHS of the above inequality increases from $\frac{J_1(\bbtheta_1)}{\bbtheta_1}$ to $\frac{J_0}{\theta_{1,0}}$. On the other hand, at $\theta_1=\bbtheta_1$, we have $\bbtheta_2=\bbtheta_1$, and the value of the RHS of the above inequality at $(\bbtheta_1,\bbtheta_2)$ is given by
\begin{align}
\text{RHS}=\frac{J_2(\bbtheta_2)}{\bbtheta_1}>\frac{J_1(\bbtheta_1)}{\bbtheta_1}.
\end{align}
As $\theta_1\to\theta_{1,0}$, or $J_1(\theta_1)\to J_0$, we know that
\begin{small}
\begin{align}
\lim_{J_1(\theta_1)\to J_0}\text{RHS}&=\lim_{J_1(\theta_1)\to J_0}-\frac{1}{\theta_1}\Big(\log\E_{z_2}\left\{e^{-\theta_2
C_2}\right\}
+\log\E_{z_1}\left\{e^{(\theta_2-\theta_1)
C_1}\right\}\Big) \nonumber\\
& = \lim_{J_1(\theta_1)\to J_0}\frac{\theta_2}{\theta_1}\Big(-\frac{1}{\theta_2}\log\E_{z_2}\left\{e^{-\theta_2
C_2}\right\}
-\frac{1}{\theta_2}\log\E_{z_1}\left\{e^{(\theta_2-\theta_1)
C_2}\right\}\Big).\label{eq:case2prooflimt}
\end{align}
\end{small}
Note further that $J_2(\theta_2)$, and hence $\theta_2$, approaches infinity as $J_1(\theta_1)\to J_0$. The first term inside the parenthesis goes to the minimum rate of the $\bR-\bD$ link, i.e., $TB\log_2(1+\tsnr_2 z_{2,\tmin})$, and the second term goes to the largest rate of the link $\bS-\bR$, i.e., $TB\log_2(1+\tsnr_1 z_{1,\tmax})$. So as long as the smallest rate of $\bR-\bD$ is less than the largest rate of the link $\bS-\bR$, the limit in (\ref{eq:case2prooflimt}) goes to $-\infty$. It is important to note that if the highest rate of $\bS-\bR$ can be supported by the link $\bR-\bD$,
 i.e.,
\begin{align}\label{eq:extremecase}
TB\log_2\left(1+\tsnr_2 z_{2,\min}\right) \ge TB\log_2(1+\tsnr_1 z_{1,\tmax}),
\end{align}
then there is no congestion at the relay node at all. In this case, $\theta_2$ can take any value greater than 0, and the only delay caused is the queue at the source. Therefore, the arrival rates are limited by the $\bS-\bR$ link, and to satisfy the statistical delay constraints, we have
\begin{align}
 R_\varepsilon(\varepsilon,D_\tmax) = \frac{J_0}{\theta_{1,0}}.
\end{align}

Now, we consider the case when (\ref{eq:extremecase}) is not satisfied. In such cases, $\theta_2\to\infty$ as $J_2(\theta_2)\to\infty$. From the continuity of the functions, we know that there must be some $(\theta_1,\theta_2)\in\Omega_\varepsilon$ such that the above inequality in (\ref{eq:case2proof1}) is satisfied with equality. Denote the smallest $\theta_1$ for such $(\theta_1,\theta_2)$ pairs as $\vvtheta_1$. Then, for all $(\theta_1,\theta_2)\in\Omega_\varepsilon$ with $\theta_1<\vvtheta_1$, (\ref{eq:case2proof1}) cannot be satisfied.

Moreover, we know as $\theta_1$ decreases, $R_1$ increases from $\frac{J_{th}(\varepsilon)}{\theta_{1,th}}$ to $\frac{J_0}{\theta_{1,0}}$. At the same time, as $\theta_2$ approaches to infinity, $R_2$ decreases from $\frac{J_{th}(\varepsilon)}{\theta_{2,th}}$ to $TB\log_2(1+\tsnr_2 z_\tmin)$. Therefore, there must be some value such that
\begin{align}\label{eq:case2proofr1r2}
R_1 =\frac{J_1(\theta_1)}{\theta_1}=R=\frac{J_2(\theta_2)}{\theta_2} = R_2
\end{align}
with the associated statistical queueing constraints denoted as $\uutheta_1$ and $\uutheta_2$, respectively. For $(\theta_1,\theta_2)\in\Omega_\varepsilon$ with $\theta_1<\uutheta_1$, we have
\begin{align}\label{eq:case2proofuu}
R_1=\frac{J_1(\theta_1)}{\theta_1}>\frac{J_2(\theta_2)}{\theta_2}=R_2.
\end{align}

\begin{figure}
  \centering
  \subfigure[$\theta_1$ and $\theta_2$ v.s. $\theta_1$.]{
    \label{fig:thetachange:a} 
    \includegraphics[width=0.3\textwidth]{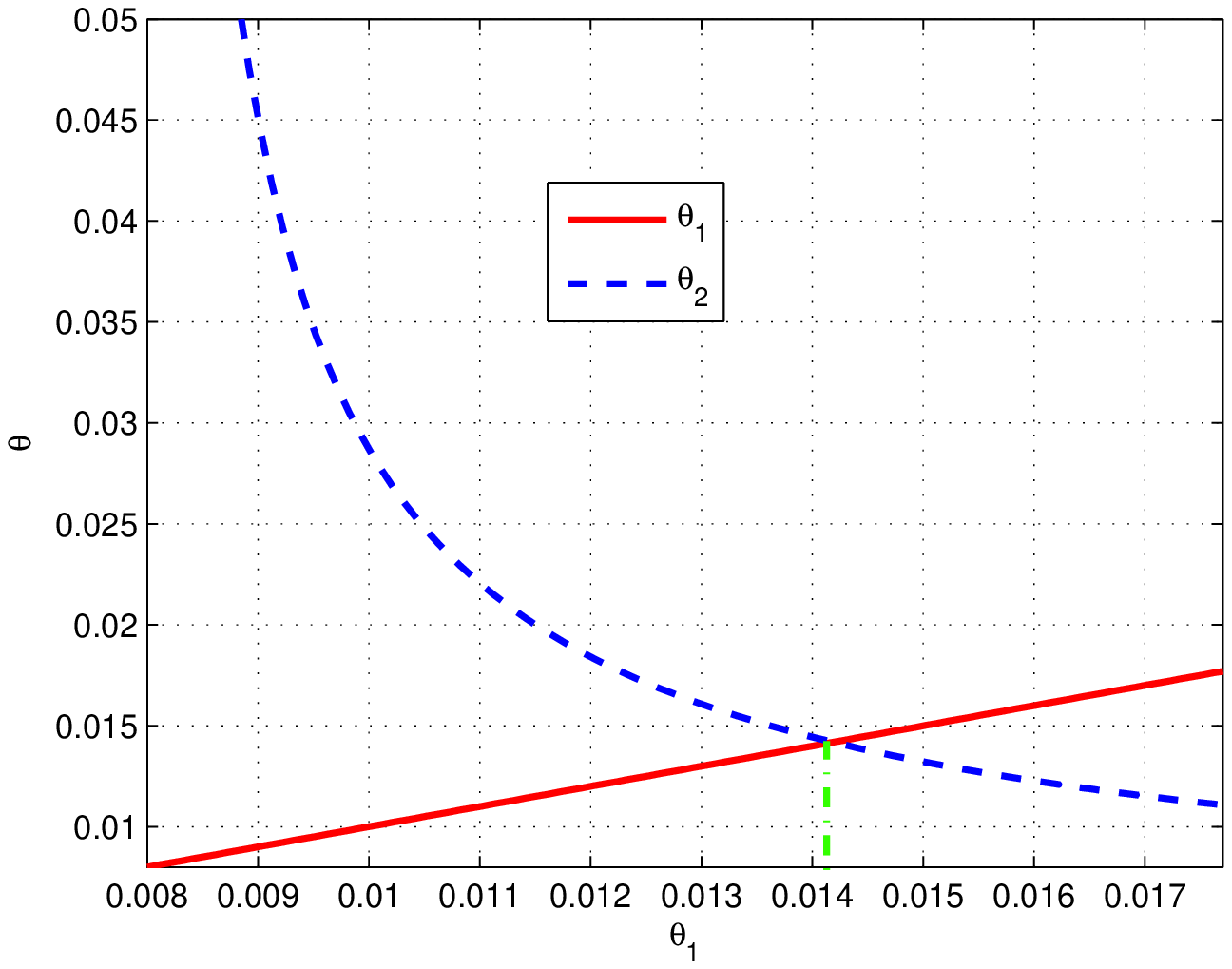}}
  \subfigure[LHS and RHS of (\ref{eq:case2proof1}) v.s. $\theta_1$.]{
    \label{fig:thetachange:b} 
    \includegraphics[width=0.3\textwidth]{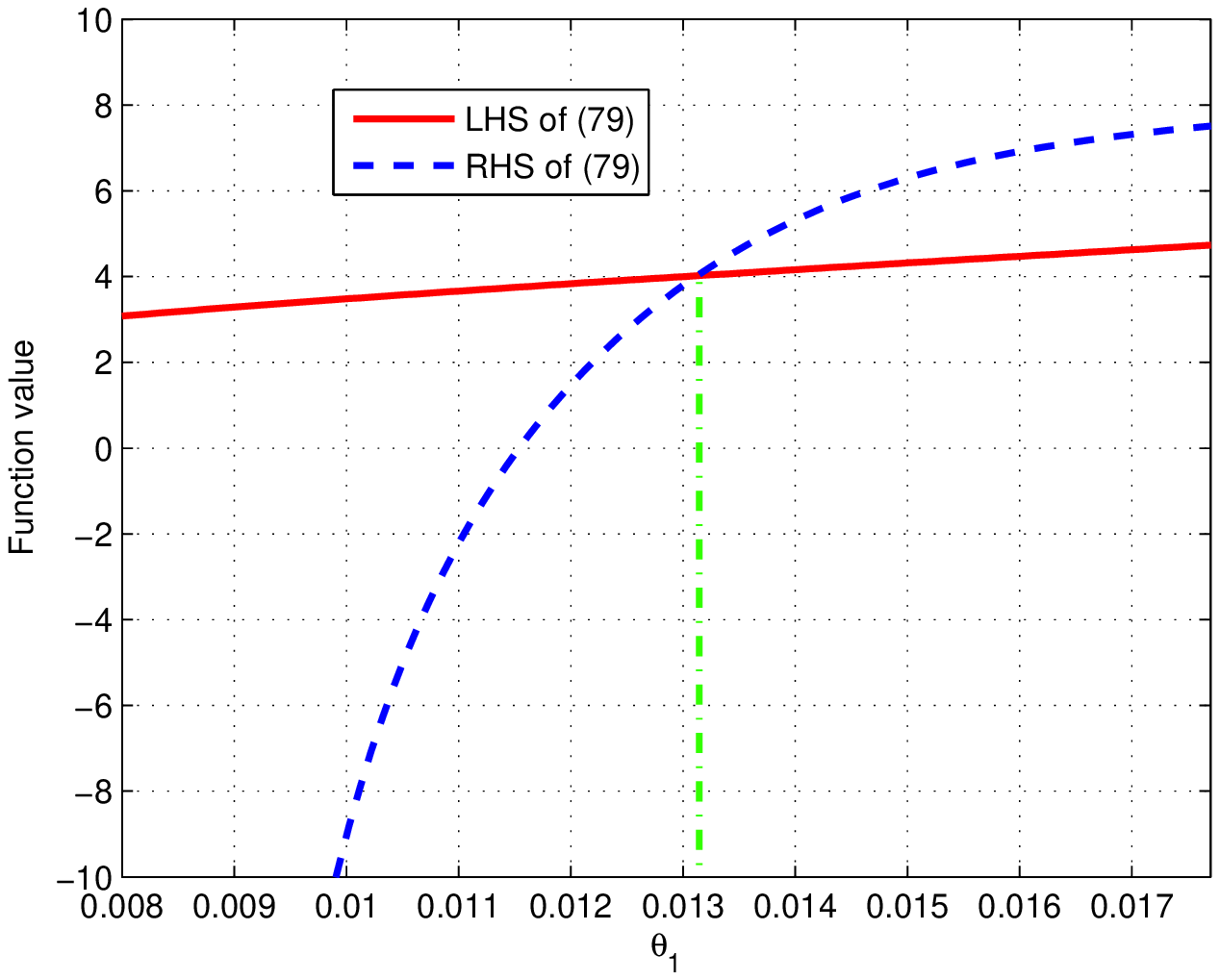}}
 \subfigure[$R_1$ and $R_2$ v.s. $\theta_1$.]{
    \label{fig:thetachange:c} 
    \includegraphics[width=0.3\textwidth]{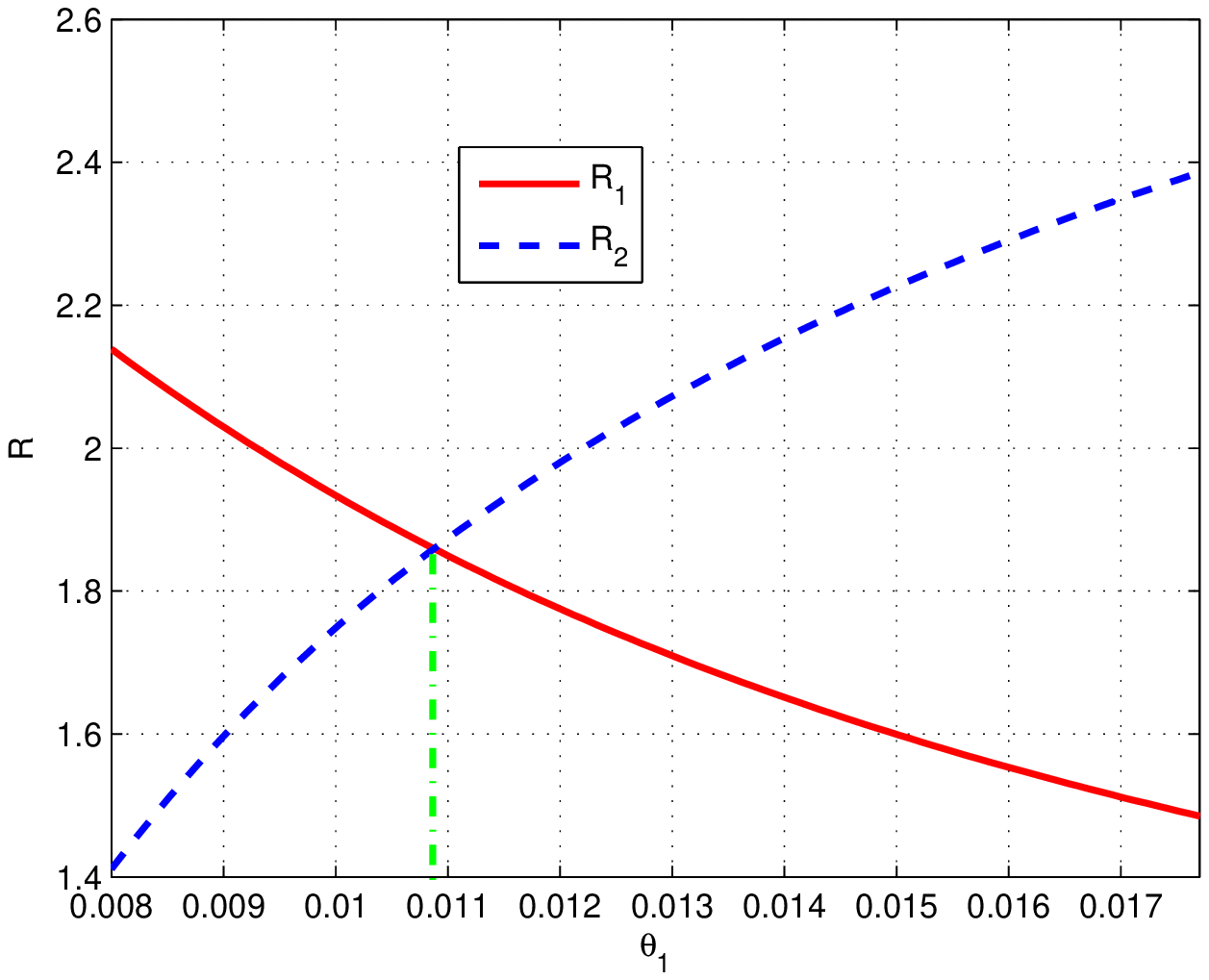}}
  \caption{The illustration of $\bbtheta_1$, $\vvtheta_1$, and $\uutheta_1$. From a)-c), the cross points give us $\bbtheta_1$, $\vvtheta_1$, and $\uutheta_1$. $\E_{z_1}\{z_1\}=\E_{z_2}\{z_2\}= 16$.}
  \label{fig:thetachange} 
\end{figure}

In the following, we can establish the comparison between $\uutheta_1$ and $\vvtheta_1$ as
\begin{align}
\uutheta_1 \le\vvtheta_1.
\end{align}
Note here that if $\frac{J_0}{\theta_{1,0}}<TB\log_2(1+\tsnr_2 z_\tmin)$, there is no $\theta_1$ for (\ref{eq:case2proofr1r2}) to be satisfied, and hence we can set $\uutheta_1$ to be 0 without affecting the following discussion based on $\vvtheta_1$, which satisfies the above claim obviously. Suppose that $\uutheta_1>\vvtheta_1$. Since at $\vvtheta_1$, the condition for \textbf{Case II} of Theorem \ref{theo:fixed} can be satisfied, we immediately see that
\begin{align}
R_E(\vvtheta_1,\vvtheta_2) = \frac{J_1(\vvtheta_1)}{\vvtheta_1}.
\end{align}
However, according to Proposition \ref{prop:upperbound} and (\ref{eq:case2proofuu}), we have
\begin{align}
R_E(\vvtheta_1,\vvtheta_2)\le\min\left\{\frac{J_1(\vvtheta_1)}{\vvtheta_1},\frac{J_2(\vvtheta_2)}{\vvtheta_2}\right\}=\frac{J_2(\vvtheta_2)}{\vvtheta_2} < \frac{J_1(\vvtheta_1)}{\vvtheta_1}
\end{align}
leading to contradiction. A numerical result provides a visualization of the aforementioned discussions on $\bbtheta_1$, $\vvtheta_1$, and $\uutheta_1$. We consider the the delay constraint given by $(\varepsilon,D_\tmax)=(0.05,1)$ in Rayleigh fading channel. We assume that $\tsnr_1 = 0$ dB, $\tsnr_2 = 3$ dB, $T=1$ ms, and $B=180$ kHz. We obtain $\theta_{1,th} = 0.0178$, and $\theta_{2,th}= 0.011$. Now, as $\theta_1$ decreases while $(\theta_1,\theta_2)\in\Omega_\varepsilon$, we plot the values of $\theta_1$ and $\theta_2$ in Fig. \ref{fig:thetachange:a}, the LHS and RHS of (\ref{eq:case2proof1}) in Fig. \ref{fig:thetachange:b}, and the values of $R_1$ and $R_2$ in Fig. \ref{fig:thetachange:c}. We can obtain $\bbtheta_1=0.0142$, $\vvtheta_1 = 0.0131$, and $\uutheta_1=0.0109$. Obviously, we can see that $\uutheta_1<\vvtheta_1<\bbtheta_1$. Note that we have $\Pr\{z_1=0\}=\Pr\{z_2=0\}=0$ for Rayleigh fading channel, and hence $J_1(\theta_2)\to\infty$ as $\theta_2\to\infty$. Note also that $z_{1,\tmax}=\infty$ and $z_{2,\tmin}=0$ for Rayleigh fading channels.

\begin{Lem1}\label{prop:maximalrateproof}
The effective capacity in this case is given by
\begin{align}\label{eq:fixresult-case2}
R_\varepsilon(\varepsilon,D_\tmax)=\sup_{(\theta_1,\theta_2)\in\Omega}R_E(\theta_1,\theta_2) = R_E(\vvtheta_1,\vvtheta_2)=\frac{J_1(\vvtheta_1)}{\vvtheta_1}.
\end{align}
\end{Lem1}

\emph{Proof:} In order to prove the proposition, we have to show that there is no other arrival rate larger than the value specified above that can be supported by the two-hop link while satisfying the statistical delay constraint. We know that for all $(\theta_1,\theta_2)\in\Omega_\varepsilon$ with $\theta_1 >\vvtheta_1$,
\begin{align}
R_E(\theta_1,\theta_2)\le\frac{J_1(\theta_1)}{\theta_1} < \frac{J_1(\vvtheta_1)}{\vvtheta_1}=R_\varepsilon(\varepsilon,D_\tmax).
\end{align}
Suppose that there exists $R>R_E(\vvtheta_{1},\vvtheta_{2})$ that can be supported by the two-hop system with $\theta_1$ and $\theta_2$. Then, $\theta_1<\vvtheta_1$. As shown above, for $(\theta_1,\theta_2)\in\Omega_\varepsilon$ with $\theta_1<\vvtheta_1$, the inequality defined in (\ref{eq:case2proof1}) cannot be satisfied, and hence $R_E(\theta_1,\theta_2)$ falls into \textbf{Case III} of Theorem \ref{theo:fixed}. In addition, with the previous characterization in (\ref{eq:bbthetafix}), we know $\theta_2 > \bbtheta_2=\bbtheta_1>\vvtheta_1$.

For \textbf{Case III.b} of Theorem \ref{theo:fixed}, i.e.,
\begin{align}
 \frac{J_2(\theta_2)}{\theta_2} <\frac{J_1(\theta_2)}{\theta_2},
\end{align}
we know that the effective capacity is decreasing in $\theta$, and as a result
\begin{align}
R_E(\theta_1,\theta_2)&=\frac{J_2(\theta_2)}{\theta_2} <\frac{J_1(\theta_2)}{\theta_2}
< \frac{J_1(\bbtheta_2)}{\bbtheta_2} =\frac{J_1(\bbtheta_1)}{\bbtheta_1}\le \frac{J_1(\vvtheta_1)}{\vvtheta_1}=R_E(\varepsilon,D_\tmax)\label{eq:case2finalproof2}
\end{align}
where $\theta_2>\bbtheta_2=\bbtheta_1>\vvtheta_1$ is incorporated.

For \textbf{Case III.a} of Theorem \ref{theo:fixed}, there exists $\tilde{\theta}_1^*\in(\theta_1,\theta_2)$ such that $\tilde{\theta}_1^*$ is the smallest solution to
\begin{small}
\begin{align*}-\frac{1}{\tilde{\theta}}\log\E_{z_1}\left\{e^{-\tilde{\theta}
C_1}\right\}
=-\frac{1}{\tilde{\theta}}\bigg(\log\E_{z_2}\left\{e^{-\theta_2
C_2}\right\}+\log\E_{z_1}\left\{e^{(\theta_2-\tilde{\theta}) C_1}\right\}\Bigg).
\end{align*}
\end{small}
With the assumption $R>R_E(\vvtheta_{1},\vvtheta_{2})$, we must have $\theta_1<\tilde{\theta}_1^*<\vvtheta_1$, and hence $J_1(\theta_1)<J_1(\tilde{\theta}_1^*)<J_1(\vvtheta_1)$. Considering the statistical delay tradeoff characterized in Lemma \ref{lemm:J1J2relation}, we must have the associated $J_2(\theta_2)>J_2(\tilde{\theta}_2^*)>J_2(\vvtheta_2)$, and hence $\theta_2>\tilde{\theta}_2^*>\vvtheta_2$. Note that with the characterizations in \cite[Lemma 2]{deli-twohop}, we can obtain the following inequality
\begin{small}
\begin{align}
-\frac{1}{\tilde{\theta}^*}\log\E_{z_1}\left\{e^{-\tilde{\theta}_1^*
C_1}\right\}& = -\frac{1}{\tilde{\theta}_1^*}\bigg(\log\E_{z_2}\left\{e^{-\theta_2
C_{2}}\right\}+\log\E_{z_1}\left\{e^{(\theta_2-\tilde{\theta}_1^*) C_1}\right\}\Bigg)\label{eq:case2proof2}\\
&<  -\frac{1}{\tilde{\theta}_1^*}\bigg(\log\E_{z_2}\left\{e^{-\tilde{\theta}^*_2
C_{2}}\right\}+\log\E_{z_1}\left\{e^{(\tilde{\theta}^*_2-\tilde{\theta}_1^*) C_1}\right\}\Bigg)
\end{align}
\end{small}
since the RHS of (\ref{eq:case2proof2}) is always greater than the LHS for all $\theta\in[0,\theta_2]$ with given $\tilde{\theta}_1^*$. That is, the condition in (\ref{eq:case2proof1}) is satisfied at $\tilde{\theta_1}^*$. This violates the definition of $\vvtheta_1$, which is the smallest solution to (\ref{eq:case2proof1}).

Combining the above discussions, we arrive at the conclusion that there is no other $\theta_1$ that can achieve higher effective capacity than (\ref{eq:fixresult-case2}). Hence, it is indeed the largest achievable constant arrival rate in this case. \hfill$\blacksquare$

The aforementioned discussions show the existence of the solution to (\ref{eq:fixresultcond}) under the statistical delay constraints. To show the uniqueness, we need the following lemma.

\begin{Lemm}
Consider the function
\begin{align}\label{eq:increasingfunction}
f(\theta_1) = J_2(\theta_2)-J_1(\theta_1)-\log\E_{z_1}\left\{e^{(\theta_2-\theta_1)C_1}\right\},\,\,\text{for}\quad \theta_1\le\bbtheta_1
\end{align}
where $(\theta_1,\theta_2)\in\Omega_\varepsilon$. If the following condition
\begin{align}\label{eq:increasingproofassumption}
\frac{dJ_2(\theta)}{d\theta}\bigg|_{\theta=\bbtheta_2}\le\frac{dJ_1(\theta)}{d\theta}\bigg|_{\theta=\bbtheta_1}
\end{align}
is satisfied, where $(\bbtheta_1,\bbtheta_2)$ is defined in (\ref{eq:bbthetafix}), then $f(\theta_1)$ is increasing in $\theta_1$.
\end{Lemm}
\emph{Proof:} Following the proof in Appendix \ref{app:J1J2relation}, we view $\theta_2$ as a function of $\theta_1$. Now taking the first derivative of $f$ over $\theta_1$, we have
\begin{small}
\begin{align}
\frac{df(\theta_1)}{d\theta_1} &= \frac{dJ_2(\theta_2)}{dJ_1(\theta_1)}\frac{dJ_1(\theta_1)}{d\theta_1}- \frac{dJ_1(\theta_1)}{d\theta_1} -\frac{\E_{z_1}\left\{e^{(\theta_2-\theta_1)C_1}C_1\right\}\left(\frac{d\theta_2}{d\theta_1}-1\right)}{\E_{z_1}\left\{e^{(\theta_2-\theta_1)C_1}\right\}}\\
& = \frac{dJ_2(\theta_2)}{dJ_1(\theta_1)}\frac{\E_{z_1}\left\{e^{-\theta_1C_1}C_1\right\}}{\E_{z_1}\left\{e^{-\theta_1C_1}\right\}} -\frac{d\theta_2}{d\theta_1}\frac{\E_{z_1}\left\{e^{(\theta_2-\theta_1)C_1}C_1\right\}}{\E_{z_1}\left\{e^{(\theta_2-\theta_1)C_1}\right\}} \nonumber\\
&\hspace{.5cm} + \frac{\E_{z_1}\left\{e^{(\theta_2-\theta_1)C_1}C_1\right\}}{\E_{z_1}\left\{e^{(\theta_2-\theta_1)C_1}\right\}} - \frac{\E_{z_1}\left\{e^{-\theta_1C_1}C_1\right\}}{\E_{z_1}\left\{e^{-\theta_1C_1}\right\}}.\label{eq:increasingproof1}
\end{align}
\end{small}
where $\frac{dJ_1(\theta_1)}{d\theta_1} = \frac{\E_{z_1}\left\{e^{-\theta_1C_1}C_1\right\}}{\E_{z_1}\left\{e^{-\theta_1C_1}\right\}}$ is substituted into (\ref{eq:increasingproof1}).

First, similar to Lemma \ref{lemm:J1}, we can show that the function $g(\theta_2)=\log\E_{z_1}\left\{e^{(\theta_2-\theta_1)C_1}\right\}$ is convex in $\theta_2$, i.e., $\frac{d^2g(\theta_2)}{d\theta_2^2}\ge0$. This tells us that the derivative of $g(\theta_2)$ is increasing in $\theta_2$, and
\begin{align}
\frac{dg(\theta_2)}{d\theta_2}\bigg|_{\theta_2=0} = \frac{\E_{z_1}\left\{e^{-\theta_1C_1}C_1\right\}}{\E_{z_1}\left\{e^{-\theta_1C_1}\right\}}.
\end{align}
Therefore,
\begin{align}
\frac{\E_{z_1}\left\{e^{(\theta_2-\theta_1)C_1}C_1\right\}}{\E_{z_1}\left\{e^{(\theta_2-\theta_1)C_1}\right\}} - \frac{\E_{z_1}\left\{e^{-\theta_1C_1}C_1\right\}}{\E_{z_1}\left\{e^{-\theta_1C_1}\right\}}\ge0.
\end{align}

Considering the definition of $(\bbtheta_1,\bbtheta_2)$ in (\ref{eq:bbthetafix}), we know that for all $(\theta_1,\theta_2)\in\Omega_\varepsilon$ with $\theta_1\le\bbtheta_1$, we have $\theta_2\ge\bbtheta_2$. Note that $J_1(\theta_1)$ and $J_2(\theta_2)$ are concave functions according to Lemma \ref{lemm:J1}, i.e., their first derivatives decreases with $\theta_1$ and $\theta_2$, respectively. Therefore, we have
\begin{small}
\begin{align}
\frac{dJ_1(\theta)}{d\theta}\bigg|_{\theta=\theta_1}&\ge \frac{dJ_1(\theta)}{d\theta}\bigg|_{\theta=\bbtheta_1},\\
\frac{dJ_2(\theta)}{d\theta}\bigg|_{\theta=\theta_2}&\le \frac{dJ_2(\theta)}{d\theta}\bigg|_{\theta=\bbtheta_2},
\end{align}
\end{small}
which, after combining with the assumption in (\ref{eq:increasingproofassumption}), gives us
\begin{align}\label{eq:increasingproof2}
\frac{dJ_1(\theta_1)}{d\theta_1}\ge\frac{dJ_2(\theta_2)}{d\theta_2}.
\end{align}

Next, recalling the statistical delay tradeoff characterized in Lemma \ref{lemm:J1J2relation}, we can see that $d\theta_2<0$ for $d\theta_1>0$, i.e., $\theta_2$ decreases as we increase $\theta_1$. Then, we can get from (\ref{eq:increasingproof2}) that
\begin{align}
\frac{d\theta_2}{d\theta_1}\le\frac{dJ_2(\theta_2)}{dJ_1(\theta_1)}.
\end{align}
Note that both $\frac{d\theta_2}{d\theta_1}$ and $\frac{dJ_2(\theta_2)}{dJ_1(\theta_1)}$ are negative values. Considering the expression in (\ref{eq:increasingproof1}), we now have
\begin{align}
\frac{df(\theta_1)}{d\theta_1} \ge \left(1-\frac{dJ_2(\theta_2)}{dJ_1(\theta_1)} \right) \left(\frac{\E_{z_1}\left\{e^{(\theta_2-\theta_1)C_1}C_1\right\}}{\E_{z_1}\left\{e^{(\theta_2-\theta_1)C_1}\right\}} - \frac{\E_{z_1}\left\{e^{-\theta_1C_1}C_1\right\}}{\E_{z_1}\left\{e^{-\theta_1C_1}\right\}}\right)\ge0.
\end{align}
That is, $f(\theta_1)$ is an increasing function in $\theta_1$. \hfill$\blacksquare$

Note that after eliminating the denominator of both sides of the equation (\ref{eq:fixresultcond}), and moving the LHS of the obtained equation to the right side, we can obtain the function given in (\ref{eq:increasingfunction}), which is increasing in $\theta_1$ for $\theta_1\le\bbtheta_1$. Therefore, the solution to the equation (\ref{eq:fixresultcond}) is unique.

\underline{\textbf{Case III}:} Assume $\theta_{1,th} < \theta_{2,th}$. For this case, at $\theta_{1,th}$, we know that
\begin{align}
R_1=\frac{J_{1}(\theta_{1,th})}{\theta_{1,th}}=\frac{J_{th}(\varepsilon)}{\theta_{1,th}}>\frac{J_{th}(\varepsilon)}{\theta_{2,th}} = \frac{J_{2}(\theta_{2,th})}{\theta_{2,th}} = R_2.
\end{align}

The queue at the relay becomes the bottleneck. We need to be careful about the effective capacity in this case. To improve the system performance, we may instead increase the queueing constraint $\theta_1$ at the source, and correspondingly, the queueing constraint $\theta_2$ at the relay can be less. Actually, relaxing the queueing constraint at the source node will not improve the performance, as will be justified later.

First, according to Lemma \ref{lemm:J1}, we can see that as $J_1(\theta_1)$ increases from $J_{th}(\varepsilon)$ to $\infty$, $\theta_1$ increases from $\theta_{1,th}$ to $\infty$. To the opposite behavior, $\theta_2$ decreases from $\theta_{2,th}$ to some finite value $\theta_{2,0}$, which is the solution to $J_2(\theta) = J_0$. Therefore, from the continuity of $\theta_2$ as a function of $\theta_1$, we again have one point $(\bbtheta_1,\bbtheta_2)\in\Omega_\varepsilon$ such that
\begin{align}\label{eq:equalthetacase3}
\bbtheta_1 = \bbtheta_2
\end{align}
and for all $\theta_1<\bbtheta_1$, we have $\theta_1<\bbtheta_1=\bbtheta_2<\theta_2$. Also, we know that $R_1$ decreases from $\frac{J_{th}(\varepsilon)}{\theta_{1,th}}$ to $TB\log_2(1+\tsnr_1 z_{1,\tmin})$, while $R_2$ increases from $\frac{J_{th}(\varepsilon)}{\theta_{1,th}}$ to some finite value $\frac{J_0}{\theta_{2,0}}$. Therefore, there must be a pair $(\theta_1,\theta_2)\in\Omega_\varepsilon$ such that
\begin{small}
\begin{align}\label{eq:case3cond}
R_1 =\frac{J_1(\theta_1)}{\theta_1}=R=\frac{J_2(\theta_2)}{\theta_2} = R_2
\end{align}
\end{small}
with the associated statistical queueing constraints denoted as $\uutheta_1$ and $\uutheta_2$, respectively. For all $\theta_1<\uutheta_1$, we have
\begin{small}
\begin{align}
R_1=\frac{J_1(\theta_1)}{\theta_1}>\frac{J_2(\theta_2)}{\theta_2}=R_2.
\end{align}
\end{small}

Note that the above result implicitly assume that $TB\log_2(1+\tsnr_1 z_{1,\tmin}) < \frac{J_0}{\theta_{2,0}}$. If this condition does not hold, then $\theta_1$ can take any value, and the only delay is introduced by the queue at the relay node. Hence, the effective capacity under the statistical delay constraint is given by
\begin{small}
\begin{align}
R_\varepsilon(\varepsilon,D_\tmax) = \frac{J_0}{\theta_{1,0}}.
\end{align}
\end{small}
Considering the queue stability condition (\ref{eq:bufferstab2}), this is possible when the average rate of $\bR-\bD$ link is larger but has more severe fading conditions.

Now, as a stark difference from the previous case, we should have
\begin{align}
\uutheta_1\ge\bbtheta_1.
\end{align}
Suppose that $\uutheta_1<\bbtheta_1$, we can show the following contradiction. First, at $\uutheta_1$, from the definition of $\bbtheta_1$ in (\ref{eq:equalthetacase3}), we have
\begin{align}
\uutheta_1<\bbtheta_1=\bbtheta_2<\uutheta_2.
\end{align}
According to the definition of $\uutheta_1$ in (\ref{eq:case3cond}), we can obtain
\begin{align}
\frac{J_1(\uutheta_1)}{\uutheta_1}=\frac{J_2(\uutheta_2)}{\uutheta_2}\Rightarrow J_1(\uutheta_1)<J_2(\uutheta_2).
\end{align}
On the other hand, according to Lemma 1, we should have
\begin{align}
J_1(\uutheta_1)>J_1(\theta_{1,th})=J_{th}(\varepsilon)=J_2(\theta_{2,th})>J_2(\uutheta_2)
\end{align}
leading to contradiction.

Since $\uutheta_1>\bbtheta_1$, with (\ref{eq:equalthetacase3}), we can see that
\begin{align}
\uutheta_1>\bbtheta_1=\bbtheta_2>\uutheta_2.
\end{align}
Now, the effective capacity $R_E(\uutheta_1,\uutheta_2)$ specializes into \textbf{Case I} of Theorem \ref{theo:fixed}, we have
\begin{align}
R_E(\uutheta_1,\uutheta_2) = \min\left\{\frac{J_1(\uutheta_1)}{\uutheta_1},\frac{J_2(\uutheta_2)}{\uutheta_2}\right\} = \frac{J_1(\uutheta_1)}{\uutheta_1}=\frac{J_2(\uutheta_2)}{\uutheta_2}.
\end{align}
Next, we can show the following result.
\begin{Lem1}
The effective capacity in this case is given by
\begin{align}\label{eq:fixresult-case3}
R_\varepsilon(\varepsilon,D_\tmax)=\sup_{(\theta_1,\theta_2)\in\Omega}R_E(\theta_1,\theta_2) = R_E(\uutheta_1,\uutheta_2)=\frac{J_2(\uutheta_2)}{\uutheta_2}=\frac{J_1(\uutheta_1)}{\uutheta_1}.
\end{align}
\end{Lem1}
\emph{Proof:} From Proposition \ref{prop:upperbound}, we know that
\begin{align}
R\le\min\left\{\frac{J_1(\theta_1)}{\theta_1},\frac{J_2(\theta_2)}{\theta_2}\right\}.
\end{align}
Now, for $\theta_1>\uutheta_1$, we can see that
\begin{align}
R_1 = \frac{J_1(\theta_1)}{\theta_1}<\frac{J_1(\uutheta_1)}{\uutheta_1}=R_\varepsilon(\varepsilon,D_\tmax)
\end{align}
and for $\theta_1<\uutheta_1$, we have $\theta_2>\uutheta_2$, and hence
\begin{align}
R_2= \frac{J_2(\theta_2)}{\theta_2}<\frac{J_2(\uutheta_2)}{\uutheta_2}=R_\varepsilon(\varepsilon,D_\tmax).
\end{align}
Therefore, $R_\varepsilon(\varepsilon,D_\tmax)$ in (\ref{eq:fixresult-case3}) is the largest achievable constant rate in this case.\hfill$\blacksquare$

\end{spacing}

\end{document}